\documentclass[letterpaper,twocolumn,10pt]{article}
\usepackage{style/usenix,epsfig,endnotes,amsmath}
\newcommand{\partitle}[1]{\smallskip \noindent \textbf{#1.}}

\newcommand{\ie}{\textit{i.e.}}

\usepackage{color}

\newcommand{\ul}[1]{\underline{#1}}

\begin{document}

\title{Towards Label-Only Membership Inference Attack against\\
Pre-trained Large Language Models}

\author{
{\rm Yu He}$^1$ \quad {\rm Boheng Li}$^2$ \quad {\rm Liu Liu}$^1$ \quad {\rm Zhongjie Ba}$^1$ \quad {\rm Wei Dong}$^2$ \quad \\ {\rm Yiming Li}$^{1,2,}$\thanks{Corresponding Author: Yiming Li.} \quad {\rm Zhan Qin}$^1$ \quad {\rm Kui Ren}$^1$ \quad {\rm Chun Chen}$^1$\\
$^1$ The State Key Laboratory of Blockchain and Data Security, Zhejiang University \\ \quad  $^2$ College of Computing and Data Science, Nanyang Technological University \\
{\tt\small \{yuherin, evence, zhongjieba, qinzhan, kuiren, chenc\}@zju.edu.cn} \\
{\tt\small \{BOHENG001, wei\_dong\}@ntu.edu.sg,}
{\tt\small liyiming.tech@gmail.com}
}

\maketitle

\begin{abstract}
Membership Inference Attacks (MIAs) aim to predict whether a data sample belongs to the model's training set or not. Although prior research has extensively explored MIAs in Large Language Models (LLMs), they typically require accessing to complete output logits (\ie, \textit{logits-based attacks}), which are usually not available in practice. In this paper, we study the vulnerability of pre-trained LLMs to MIAs in the \textit{label-only setting}, where the adversary can only access generated tokens (text). We first reveal that existing label-only MIAs have minor effects in attacking pre-trained LLMs, although they are highly effective in inferring fine-tuning datasets used for personalized LLMs. We find that their failure stems from two main reasons, including better generalization and overly coarse perturbation. Specifically, due to the extensive pre-training corpora and exposing each sample only a few times, LLMs exhibit minimal robustness differences between members and non-members. This makes token-level perturbations too coarse to capture such differences. 

To alleviate these problems, we propose \textbf{PETAL}: a label-only membership inference attack based on \textbf{PE}r-\textbf{T}oken sem\textbf{A}ntic simi\textbf{L}arity. Specifically, PETAL leverages token-level semantic similarity to approximate output probabilities and subsequently calculate the perplexity. It finally exposes membership based on the common assumption that members are `better' memorized and have smaller perplexity. We conduct extensive experiments on the WikiMIA benchmark and the more challenging MIMIR benchmark. Empirically, our PETAL performs better than the extensions of existing label-only attacks against personalized LLMs and even on par with other advanced logit-based attacks across all metrics on five prevalent open-source LLMs\footnote{Code is available at \href{https://zenodo.org/records/14725819}{https://zenodo.org/records/14725819}.}. Our study highlights that pre-trained LLMs remain vulnerable to privacy risks posed by MIAs even in the most challenging and realistic setting, calling for attention to develop more effective defenses.
\end{abstract}

\section{Introduction}
Large Language Models (LLMs) which use very large architectures and are pre-trained on massive corpora have the capability to generate high-quality and human-like responses. They can handle various complex downstream tasks such as text summarization, code generation, and semantic analysis \cite{hoang2019efficient, vaithilingam2022expectation, gilbert2023large} even without changing any model parameters \cite{brown2020language}. It is foreseeable that LLMs will change the way of human creation profoundly \cite{achiam2023gpt,reid2024gemini,bai2023qwen}.

However, with the deployment of LLMs comes the necessity to address the associated privacy risks \cite{ishihara2023training}. For instance, existing studies demonstrate that LLMs memorize training data well and personally identifiable information can be extracted from their generation (i.e., extraction attacks \cite{carlini2021extracting,lee2023language}). Another common source of privacy risk stems from membership inference attacks (MIAs \cite{shokri2017membership}). For a given target model, MIAs enable adversaries to determine whether a sample is present in its training data or not. Beyond use for constructing extraction attacks and evaluating data leakages, MIAs can also audit machine unlearning \cite{ginart2019making,bourtoule2021machine} and detect whether an LLM has used one's pieces unauthorizedly for its pre-training \cite{shi2023detecting}. 

Existing MIAs against language models mainly assume a logits-based setting. Since the model fits members better than non-members \cite{carlini2021extracting}, they can calculate a membership score (e.g., the cross-entropy loss or perplexity) by having the direct access to the complete logits (probabilities) of generated token(s). Unfortunately, many public commercial LLM-APIs only provide users with generated tokens (texts) rather than complete logits, which makes previous attacks largely inapplicable. These limitations inspire us to raise an important question: \textit{can we infer membership by checking generated tokens only, without accessing the complete logits (i.e., label-only setting)?} This setting is more practical yet more challenging: the tokens reveal only limited information than logits, making membership scores difficult to calculate.

We focus on memorization in pre-training because this phase is more prevalent and presents greater technical challenges (see Section \ref{sec: 3.1} for a detailed analysis). We first summarize existing MIAs designed for language models and evaluate their performances on pre-trained LLMs. Specifically, we identify a fundamental limitation of label-only setting based MIAs on the pre-training phase of LLMs, i.e., attacks based on the robustness gap between members and non-members are largely ineffective. This is because better generalization of pre-trained LLMs makes perturbations at the token level too coarse to capture the tiny gap in distance to the decision boundary between members and non-members.

To mitigate this gap, we intend to design a label-only MIA based on the properties of LLMs. Our motivation stems from an intriguing observation that given preceding tokens, the probability of a particular token being generated by the target model is potentially correlated with the semantic similarity between that token and the token actually generated by the target model (use preceding tokens as input). 
As such, we can leverage token-level semantic similarity to approximate output probabilities and subsequently calculate the perplexity (loss) of samples. Specifically, we use univariate linear regression to map token-level semantic similarity to output probabilities. However, it is difficult to obtain regression parameters since adversaries cannot access the output probabilities from the target model. Inspired by the concept of shadow models \cite{shokri2017membership}, we use an open-source LLM as the surrogate model to perform pre-regression and obtain the required parameters for estimating the output probabilities of the target samples. We note that although there are differences between the surrogate model and the target model, this substitution works because the main role of the surrogate model is to align the distribution of semantic similarity scores with the true output probabilities, which will be detailed in Section \ref{sec: 4.3}. Finally, following the perspective that members are memorized ``better'' and have smaller losses \cite{yeom2018privacy}, the adversary can easily identify members by thresholding the approximated perplexity. 

Based on that, we propose \textbf{PETAL}, a membership inference attack leveraging \textbf{PE}r-\textbf{T}oken sem\textbf{A}ntic simi\textbf{L}arity to explore memorization of pre-trained LLMs in real-world settings. For any target sample, our PETAL executes in four stages: \textbf{1)} Obtain semantic similarity-probability pairs for all tokens on the surrogate model. \textbf{2)} Perform a univariate linear regression on the two variables. \textbf{3)} Query the target model to get token-level semantic similarity. \textbf{4)} Approximate the probabilities assigned by the target model to each token to calculate an alternative perplexity and attack by thresholding.

We conduct experiments on five prevalent open-source LLMs to evaluate the performance of PETAL, with comparisons to existing label-only attacks and other advanced logit-based attacks. Results on two classical benchmarks, including WikiMIA \cite{shi2023detecting} and MIMIR \cite{duan2024membership}, demonstrate that our PETAL not only significantly outperforms existing label-only attacks but is also comparable to the logits-based attacks requiring more capacities. We further extensively study all parameters that could impact the performance of PETAL, such as the query budget, the granularity of output probability estimation, the selection of the surrogate model, as well as the length of input text, the decoding strategy, and the weighting strategy. We also include two case studies on fine-tuned LLMs to demonstrate the robustness of PETAL. Following that, we conduct a detailed analysis of several well-established defense strategies and their mitigating effects against PETAL (and other attacks). We also note that existing MIA research lacks exploration of closed-source LLMs, so we conduct two case studies on Gemini-1.5-Flash \cite{team2023gemini} and GPT-3.5-Turbo-Instruct \cite{achiam2023gpt}. 

Our main contributions are summarized as follows:
\begin{itemize}
\setlength\itemsep{-1pt}
\item To the best of our knowledge, we are the first to explore whether pre-trained LLMs suffer from membership inference attacks under the label-only setting.
\item We reveal that label-only MIAs based on the robustness gap are largely ineffective in the pre-training phase of LLMs due to potential reasons like better generalization and overly coarse perturbation.
\item We propose PETAL, a simple yet effective label-only MIA against the pre-training phase of LLMs. PETAL leverages token-level semantic similarity to approximate output probabilities for membership inference. 
\item Extensive evaluations on two benchmarks and five open-source pre-trained LLMs show that PETAL is on par with existing logits-based attacks across various metrics.
\end{itemize}

\section{Background \& Related Work}
In this section, we introduce the relevant background on pre-trained LLMs as well as membership inference attacks.

\subsection{Pre-trained Language Models}
Language models encapsulate a likelihood distribution across token sequences. In this paper, we focus on autoregressive language modeling \cite{bengio2000neural,mikolov2010recurrent,radford2018improving}, since this method has become the predominant force within large-scale language models \cite{ouyang2022training,wang2021gpt,touvron2023llamaa,radford2019language}. Today's most powerful LLMs such as GPT-4 \cite{achiam2023gpt} and Gemini-1.5 \cite{reid2024gemini}, all follow this paradigm. An autoregressive language model generates the probability distribution for the next token given preceding tokens. Concretely, we let \(f_\theta(t_i\mid t_1,\ldots,t_{i-1})\) denote the likelihood  \(\mathbf{Pr}(t_i\mid t_1,\ldots,t_{i-1})\) of token \(t_i\) when evaluating the language model \(f\) with parameters \(\theta\), where \(t_1,...,t_{i}\) is a sequence of tokens from a vocabulary \(\mathcal{V}\). Hence, the distribution \(\mathbf{Pr}(t_1,\ldots,t_n)\) can be derived by employing the chain rule of probability:
\begin{equation}
    \mathbf{Pr}(t_1,\ldots,t_n)=\Pi_{i=1}^n\mathbf{Pr}(t_i\mid t_1,\ldots,t_{i-1}).
\end{equation}
The language model can be thus optimized by minimizing the negative log-probability of each data point within the massive pre-training sets. Formally, the loss function can be defined as follows:
\begin{equation}
    \mathcal{L}(\theta)=-\log\Pi_{i=1}^n\mathbf{Pr}(t_i\mid t_1,\ldots,t_{i-1}).
\end{equation}
This setting can be qualitatively regarded as memorizing the tokens following each specific prefix in the training set. It is worth noting that current state-of-the-art LLMs are pre-trained on extensive corpora and expose each sample only once or a few times (fewer than three) \cite{touvron2023llamaa}, leading to nearly consistent loss on both the training and test sets \cite{brown2020language}.

In this paper, we primarily focus on the label-only settings, hence requiring a careful consideration of specific text generation strategies. A language model can generate new text (potentially conditioned on given prefix \(t_1,...,t_{i}\)) by iteratively sampling \(\hat{t}_{i+1}\sim f_\theta(t_{i+1}\mid t_1,\ldots,t_{i})\) and subsequently using \(\hat{t}_{i+1}\) as input to sample \(\hat{t}_{i+2}\sim f_\theta(t_{i+2}\mid t_1,\ldots,t_{i+1})\). This process repeats until the designated stopping criterion is satisfied. Existing generating strategies can be categorized into deterministic methods such as beam search and contrastive search (selects a sample from the set of most probable candidates while ensuring the generated output remains discriminative enough concerning the preceding context) \cite{su2022a}, and stochastic methods like nucleus sampling (draws a sample from the smallest vocabulary subset with total probability mass above a given threshold) \cite{holtzman2019curious}. 

\vspace{-0.2em}
\subsection{Membership Inference Attacks}
Membership Inference Attacks, initially introduced by Shokri et al.\cite{shokri2017membership}, continue to be an intriguing topic within the ML community. Depending on the attacker's accessibility to the target model, MIAs can be classified into black-box attacks (assume the access to output logits) \cite{shokri2017membership,salem2018ml,yeom2018privacy,watson2021importance,carlini2022membership,ye2022enhanced,liu2022membership,he2024difficulty} and white-box attacks (assume the access to the model parameters) \cite{sablayrolles2019white, nasr2019comprehensive,leino2020stolen}. Choquette et al. \cite{choquette2021label} attempt to perform MIAs under stricter conditions (label-only) and showcase that logits are not imperative. 


Specifically in NLP, most studies on MIA issues \cite{carlini2021extracting,mireshghallah2022quantifying,mireshghallah2022empirical,mattern-etal-2023-membership,fu2023practical,shi2023detecting} adhere to a black-box setting.  

\partitle{Focusing on the Pre-training Phase} 
Carlini et al. \cite{carlini2021extracting} predict an example to be a member if its loss (i.e., perplexity) is less than a certain threshold. Mireshghallah et al. \cite{mireshghallah2022quantifying} share a similar loss-based idea and further incorporate with a public model (reference model) to estimate sample hardness \cite{carlini2022membership}. Shi et al. \cite{shi2023detecting} introduce MIN-K\% PROB based on the hypothesis that an unseen example is likely to contain a few outlier words with low probabilities. Duan et al. \cite{duan2024membership} argue that existing benchmarks have a temporal distribution shift between members and non-members, which could potentially improve attack performances. They create a more challenging benchmark (i.e., MIMIR), where the training and non-training texts are maximally similar to each other.

\partitle{Focusing on the Fine-tuning Phase}
Both Mireshghallah et al. \cite{mireshghallah2022empirical} and Fu et al. \cite{fu2023practical} use reference models, with the former employing a public pre-trained model that is not fine-tuned, and the latter utilizing a model re-fine-tuned on another similar dataset. Mattern et al. \cite{mattern-etal-2023-membership} propose a neighborhood attack that leverages neighborhood samples as an alternative to reference models, since the assumption of access to sufficient samples closely resembling the original training data is strong and arguably unrealistic. 

One limitation of these studies is that many web services employing LLMs restrict users to viewing tokens generated by the model solely instead of output logits. Though calls for releasing logits for transparency have been made \cite{bucknall2023structured}, model creators like OpenAI\footnote{\url{https://platform.openai.com/docs/api-reference/completions/create}} typically do not release full logits or may limit the output to the top-$k$ logits for each query, which is still insufficient for current logits-based attacks. Notably, the attack proposed by \cite{tang2023assessing} is probably closest to our work as it tries to achieve label-only MIAs targeted at the fine-tuning phase of Seq2Seq models (e.g., Bart \cite{lewis2020bart}). However, their robustness-based attack proves ineffective against pre-trained LLMs due to better generalization and overly coarse perturbation (see Section~\ref{sec: 3.2} \& \ref{sec: 3.3} for a detailed analysis). To the best of our knowledge, we are the first to implement and evaluate effective \textbf{label-only} membership inference attacks targeted at \textbf{the pre-training phase of LLMs}.

\begin{table*}
    \centering
    \caption{Taxonomy of MIAs against LMs over existing works. Here we use ``robustness + [method]'' to denote different variants of Tang et al.'s attack, utilizing [method] (e.g., Word Substitution) to estimate robustness. We bold our results when they surpass all label-only attacks and underline them when they are comparable to logits-based attacks (i.e., surpassing at least two logits-based attacks). The results demonstrate that logits-based attacks largely outperform existing label-only attacks when targeting LLMs.}
    \vspace{-2mm}
        \scalebox{0.86}
        {
        \begin{tabular}{clccccc} \toprule
         Attack Settings & \multicolumn{1}{l}{Method} & Original Target Phase & Reference & AUC $\uparrow$ & Balanced Accuracy $\uparrow$ & TPR@1\%FPR  $\uparrow$\\
         \cmidrule(lr){1-2} \cmidrule(lr){3-3} \cmidrule(lr){4-4} \cmidrule(lr){5-5} \cmidrule(lr){6-6} \cmidrule(lr){7-7}
         - & Random Guess & - & \XSolidBrush & 0.50 & 0.50 & 1.0\%\\
         \midrule
        \multirow{5}{*}{\textbf{Logits-based}} & PPL attack \cite{carlini2021extracting,yeom2018privacy}& Pre-training & \XSolidBrush & 0.64 & 0.62 & 6.2\%\\
         & reference attack \cite{carlini2021extracting}& Pre-training & \CheckmarkBold & 0.50 & 0.52 & 2.1\%\\
         & zlib attack \cite{carlini2021extracting}& Pre-training & \CheckmarkBold & 0.64 & 0.62 & 4.9\%\\
         & neighborhood attack \cite{mattern-etal-2023-membership}& Fine-tuning & \CheckmarkBold & 0.66 & 0.62 & 0.8\%\\
         & MIN-K\% PROB \cite{shi2023detecting} & Pre-training & \XSolidBrush & 0.66 & 0.63 & 8.8\%\\
         \midrule
         \multirow{4}{*}{\textbf{Label-only}} & robustness + WS \cite{tang2023assessing} & Fine-tuning  & \XSolidBrush & 0.49 & 0.52 & 0.3\%\\
         & robustness + RS \cite{tang2023assessing} & Fine-tuning & \XSolidBrush & 0.50 & 0.51 & 0.3\%\\
         & robustness + BT \cite{tang2023assessing} & Fine-tuning & \XSolidBrush & 0.51 & 0.53 & 1.0\%\\
         & PETAL (Ours) & Pre-training & \XSolidBrush & \textbf{\ul{0.64}} & \textbf{\ul{0.62}} & \textbf{\ul{4.9\%}}\\
         \bottomrule
         \end{tabular}
         }
         \begin{tablenotes}
            \centering
            \item \CheckmarkBold: reference-based $\,\,\,\,\,\,\,\,\,\,\,\,\,\,\,\,\,\,\,\,\,\,\,\,\,\,\,\,\,\,\,\,\,\,\,\,\,\,\,\,\,\,\,\,\,\,\,\,\,\,\,\,\,\,\,\,\,\,\,\,\,\,\,\,\,\,\,\,\,\,\,\,\,\,\,\,\,\,\,\,\,\,\,\,\,\,\,\,\,\,$ \XSolidBrush: reference-free 
         \end{tablenotes}
    \label{tab: summary}
\vspace{-1.2em}
\end{table*}

\vspace{-0.3em}
\section{Revisiting Label-only MIAs}

In this section, we formally define membership inference attacks and investigate whether label-only MIAs designed for the fine-tuning phase of language models \cite{tang2023assessing} work for pre-trained LLMs.

\subsection{Preliminaries}
\label{sec: 3.1}
MIAs aim to infer whether a target sample belongs to the training set of a given trained model. We follow the initial definition of MIAs by \cite{shokri2017membership}. Given an LLM \(f_\theta\) parameterized by weights \(\theta\) and its pre-training dataset \(\mathcal{D}\) sampled from the underlying distribution \(\pi\), \(f_\theta\) is solely pre-trained on \(\mathcal{D}\). Given any target point \(x_i\in\pi\), the adversary would use an attack method \(\mathcal{A}\) to predict its membership state \(m_i\), where \(m_i = 1\) if \(x_i\in\mathcal{D}\); otherwise \(m_i = 0\). We can thus make a formal definition of \(\mathcal{A}\) as follows:
\begin{equation}
\mathcal{A}(x_i,\theta)=\mathds{1}[\mathbf{Pr}(m_i=1\mid x_i,\theta)\geq\tau],
\end{equation}
where \(\tau\) indicates a threshold used for decision-making. The adversary does not have any knowledge of \(\mathcal{D}\). 

\partitle{Adversary’s Capabilities} In a label-only setting, the adversary's information is limited to only the sequence of generated new tokens for any input. For a more realistic assumption, he has no access to the structure of \(f_\theta\) or a sufficient number of samples drawn from \(\pi\setminus\mathcal{D}\) to train certain shadow/reference models \cite{watson2021importance,carlini2022membership}, as such information is not always available (e.g., not released by owners like OpenAI \cite{openai2023gpt4}). Stepping back, even with such access, the computational overhead of training one or multiple similar models on the incredible scale of pre-training data would be unaffordable for common attackers. Thus, this presents more challenging obstacles compared to attacking the fine-tuning phase\textemdash the adversary cannot carefully craft a reference model similar to the target model to estimate sample hardness \cite{watson2021importance}.

\partitle{Metrics} We follow prior works \cite{duan2024membership, shi2023detecting, mattern-etal-2023-membership} to use three convincing metrics to measure the performance of MIAs:
\begin{itemize}\setlength{\itemsep}{0.1em} \setlength{\parskip}{0.1em}
\item \textbf{Balanced Accuracy.} The simplest method to evaluate attack efficacy that measures how often an attack correctly predicts membership on a balanced dataset of members and non-members~\cite{choquette2021label, shokri2017membership, sablayrolles2019white, yeom2018privacy}.
\item \textbf{AUC.} The most commonly used method to interpret the Receiver Operating Characteristic (ROC) curve~\cite{sankararaman2009genomic} is by calculating the area under the curve (AUC). It reflects the average-case success of membership inference.
\item \textbf{TPR at Low FPR.} The latest metric used to evaluate attack efficacy focuses on the TPR of the attack when the threshold \(\tau\) is set to a large value to achieve an extremely low FPR. This metric is recommended in classical MIAs against image classification models since it directly reflects the actual extent of privacy leakage of the model towards its training samples ~\cite{carlini2022membership}. 
\end{itemize}

It is worth noting that existing studies on MIAs against pre-trained LLMs primarily report AUC results \cite{shi2023detecting, duan2024membership, zhang2024min, xie2024recall}. This can be attributed to the focus on large-scale extraction of pre-training data in this setting \cite{carlini2021extracting}. Additionally, in scenarios such as auditing machine unlearning \cite{yao2024machine} or detecting copyright infringement, the AUC is more important \cite{shi2023detecting}.

\subsection{Robustness-based Label-only MIA}
\label{sec: 3.2}

Following a max-margin perspective, non-member data points typically exhibit lower robustness. That is, a model's predictions for non-members are more easily altered by input perturbations compared to those for members \cite{tanay2016boundary,hu2019new}. This is mainly because non-members are closer to the decision boundary. Based on this, Choquette et al. \cite{choquette2021label} have designed the first label-only attack which is as-effective as attacks requiring access to logits for classification models. Tang et al. \cite{tang2023assessing} also exploit robustness as potential MIA signals and evaluate its effectiveness in the fine-tuning phase of language models. Given any target sample, their method selects its preceding \(n\%\) tokens as original input and creates additional augmented inputs via different text perturbation strategies. The attacker then uses the collected inputs to prompt the model to generate the original output and augmented outputs (with a maximum length equal to that of the original sample). Finally, he uses the text similarity scores between all outputs (i.e., original and augmented outputs) with the ground truth (i.e., the remaining tokens) as the feature to identify members.

In our experiment, we mainly consider three text perturbation methods: Random Swapping (RS), Back Translation (BT) and Word
Substitution (WS). RS randomly chooses 15\% of words in a sentence and swaps them with adjacent words. BT translates the source language to another language and translates it back to the source. WS randomly masks out 15\% words in a sentence, then employs Bert-large \cite{kenton2019bert} to fill in and generate similar sentences. For each sample, we generate three augmented inputs.

\subsection{Evaluation and Analysis}
\label{sec: 3.3}

Table \ref{tab: summary} presents our summary about MIAs designed for language models. In addition to robustness-based attacks, we consider five advanced logits-based attacks:
\begin{itemize}\setlength{\itemsep}{0.1em} \setlength{\parskip}{0.1em}
\item \textbf{PPL Attack \cite{carlini2021extracting,yeom2018privacy}} directly takes the perplexity \(\mathcal{P}\) of the target sample \(x\) on the target model \(f_\theta\) as the membership score for detection: \(\mathcal{A}(x,\theta)=\mathds{1}[-\mathcal{P}(x,f_\theta)\geq\tau]\).
\item \textbf{Reference Attack \cite{carlini2021extracting}} accounts for the hardness of
the sample by obtaining its perplexity from another reference model: \(\mathcal{A}(x,\theta)=\mathds{1}[\mathcal{P}(x,f_{ref})-\mathcal{P}(x,f_\theta)\geq\tau]\).
\item \textbf{Zlib Attack \cite{carlini2021extracting}} accounts for the hardness of
the target sample by using its zlib compression size: \(\mathcal{A}(x,\theta)=\mathds{1}[-\frac{\mathcal{P}(x,f_\theta)}{\mathrm{zlib}(x)}\geq\tau]\).
\item \textbf{Neighborhood Attack \cite{mattern-etal-2023-membership}} accounts for the hardness of
the target sample by comparing its original membership score with scores of synthetically generated neighbor texts \(\Tilde{x}_i\): \(\mathcal{A}(x,\theta)=\mathds{1}[\frac1n\sum_{i=1}^n\mathcal{P}(\tilde{x},f_\theta)-\mathcal{P}(x,f_\theta)\geq\tau]\). 
\item  \textbf{MIN-K\% PROB \cite{shi2023detecting}} uses the k\% of tokens with the lowest likelihoods to compute membership score instead of averaging over all tokens as in perplexity: \(\mathcal{A}(x,\theta)=\mathds{1}[\frac{1}{\mid\mathrm{Min-K}(t)\mid}\sum_{t_{i}\in\mathrm{Min-K}(t)}\log p(t_{i}|t_{1},\ldots,t_{i-1})\geq\tau]\).
\end{itemize}

For the model and dataset settings, we employ Pythia-6.9B \cite{biderman2023pythia} and WikiMIA \cite{shi2023detecting} here. For decoding strategy, we utilize greedy search here and leave a more detailed analysis of decoding strategy in Section~\ref{sec: 5.3}.

Contrary to intuition, the results in Table \ref{tab: summary} indicate that robustness-based attacks are significantly weaker than logits-based attacks across all metrics (nearly equivalent to random guessing), which implies that \textbf{robustness estimated from perturbations probably can not serve as a reliable indicator for inferring membership}. This can be attributed to the fact that LLMs are pre-trained on extensive corpora and expose each sample only once or a few times (fewer than three \cite{touvron2023llamaa}), leading to superior generalization compared to classification models and fine-tuned language models. Thus, the gap in distance between members and non-members to the decision boundary decreases. However, the adversary can only perturb input prefixes at the token level, which is too coarse to accurately capture this gap for distinguishing members/non-members. We believe that finer-grained perturbations (e.g., embedding-level) could better capture robustness differences between members and non-members. However, this requires access to the parameters of model embedding layers, which does not align with our threat model. Future work could be conducted to investigate the specific impact of perturbation granularity on robustness-based attacks, which is out of the scope of this paper. To address the gap in label-only MIAs against pre-trained LLMs, we consider designing more effective attacks by exploiting the specific properties of LLMs.

\vspace{-0.5em}
\section{Methodology}
\label{sec: 4}
\begin{figure}[t]
    \centering
    \includegraphics[width=0.473\textwidth]{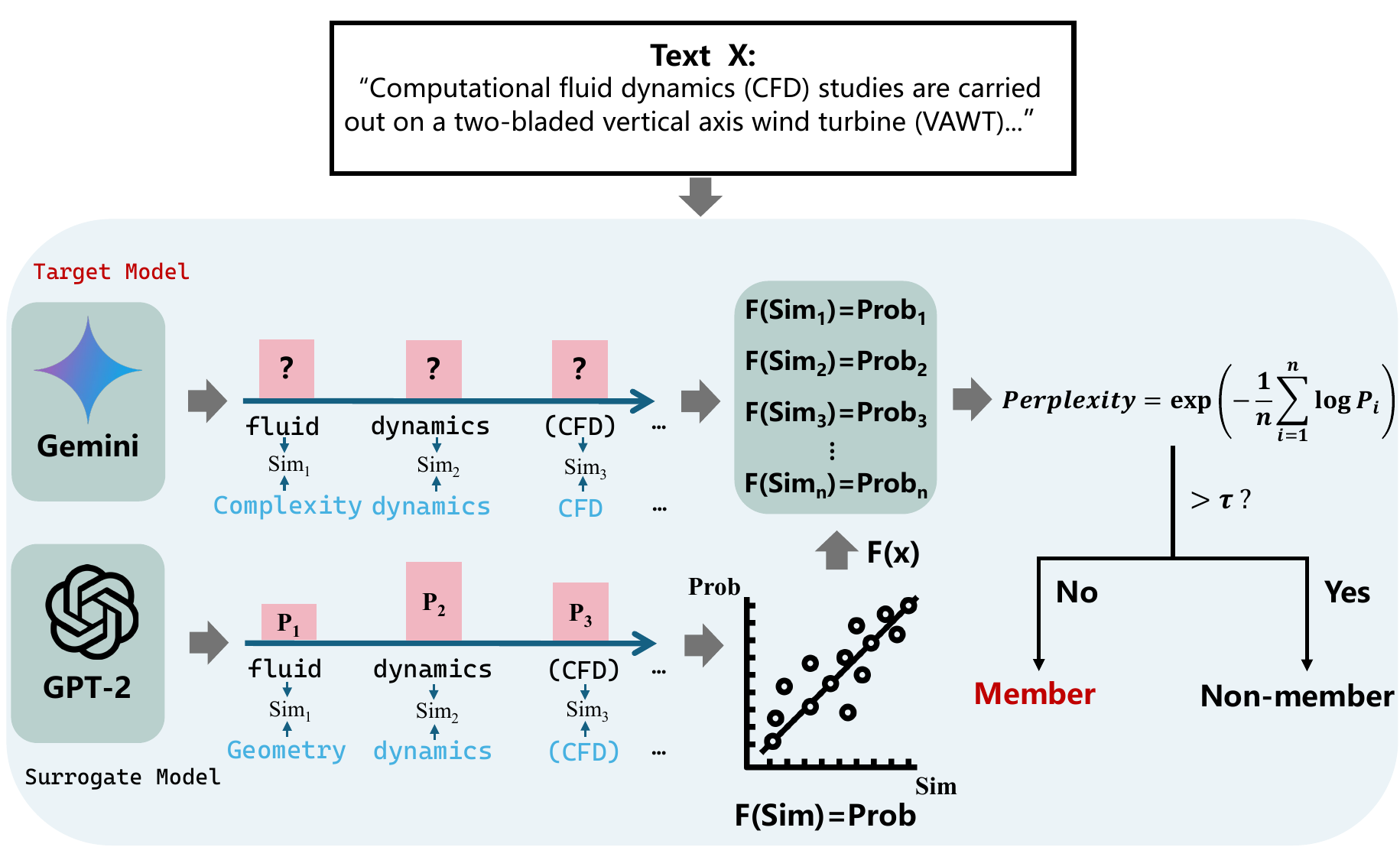}
    \caption{General attack pipeline of our PETAL.}
    \label{fig: pipeline}
\vspace{-1.5em}
\end{figure}

In this section, we propose the first effective label-only MIA designed for pre-trained LLMs. We start by introducing a novel insight: only the first generated token contains effective MIA signals, which determines the object for extracting membership information. Then, we detail how to use the semantic similarity between the first generated token and a particular token to approximate the latter's output probability. Finally, we explain how to identify members using the approximated output probabilities and why our method works. The detailed pipeline of our PETAL is illustrated in Figure \ref{fig: pipeline}.

\subsection{Focusing on the First Generated Token}
\label{sec: 4.1}
Recall that LLMs are trained to minimize the negative log-probability of each sample within the training set. Thus, a natural assumption regarding membership inference is that, given the first part of a sequence from the member set as input, LLMs can more accurately ``predict'' the remaining part of that sequence compared to a sequence from the non-member set. However, we argue that \textbf{in a single query only the first generated token contains effective MIA signals} because prediction errors accumulate throughout the process of iteratively sampling. The following toy example illustrates how this leads to a degradation in the differentiation between member and non-member sets. 

\partitle{Example 1} 
Assume a sequence from the member set is ``Donald Trump is a president''. However, when feeding ``Donald'' into the model, it might incorrectly predict the next token as ``Duck'' and iteratively output ``Donald Duck is a cartoon character''. This causes a member to be misclassified as a non-member. On the contrary, if we input ``Donald'', ``Donald Trump'', …, ``Donald Trump is a'' sequentially and only focus on the first generated token, even if ``Trump'' is predicted as ``Duck'', errors won't accumulate and the target sequence would most likely be inferred as a member. 

\begin{figure}[t]
    \centering
    \includegraphics[width=0.44\textwidth]{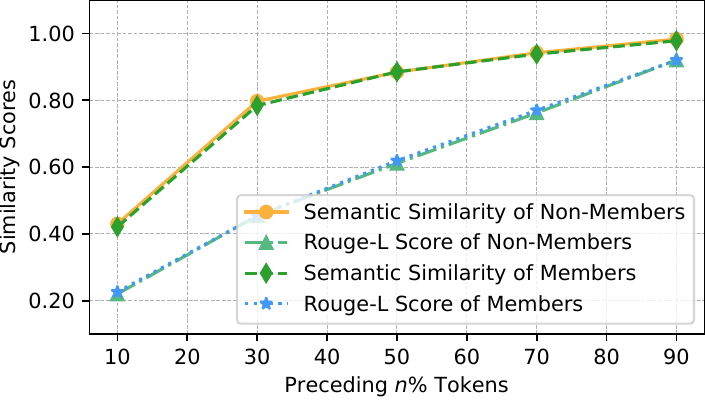}
    \caption{The similarity scores distribution of members and non-members. There is no notable gap in the scores between members and non-members.}
    \label{Figure: only_one_attack}
\vspace{-1.5em}
\end{figure}

To further verify this, we select the first $n\%$ tokens of each sample from WikiMIA dataset \cite{shi2023detecting} as inputs to Pythia-2.8B model \cite{biderman2023pythia} and calculate the similarity between the generation and the ground truth, including semantic similarity \cite{reimers2019sentence} and ROUGE-L scores \cite{lin2004looking}. Figure \ref{Figure: only_one_attack} demonstrates that language models will not show distinct bias toward the member samples if the sampling process is iterated. In other words, only the first generated token directly reflects the model's bias and contains effective MIA signals, since its corresponding prefix does not include the iteratively generated tokens. We will further verify this in the ablation study in Section \ref{sec: 5.3}.

\begin{figure}[t]
    \centering
    \subfloat[Text: \textcolor{red}{``The 67th British…''}]{\includegraphics[width=1.7in]{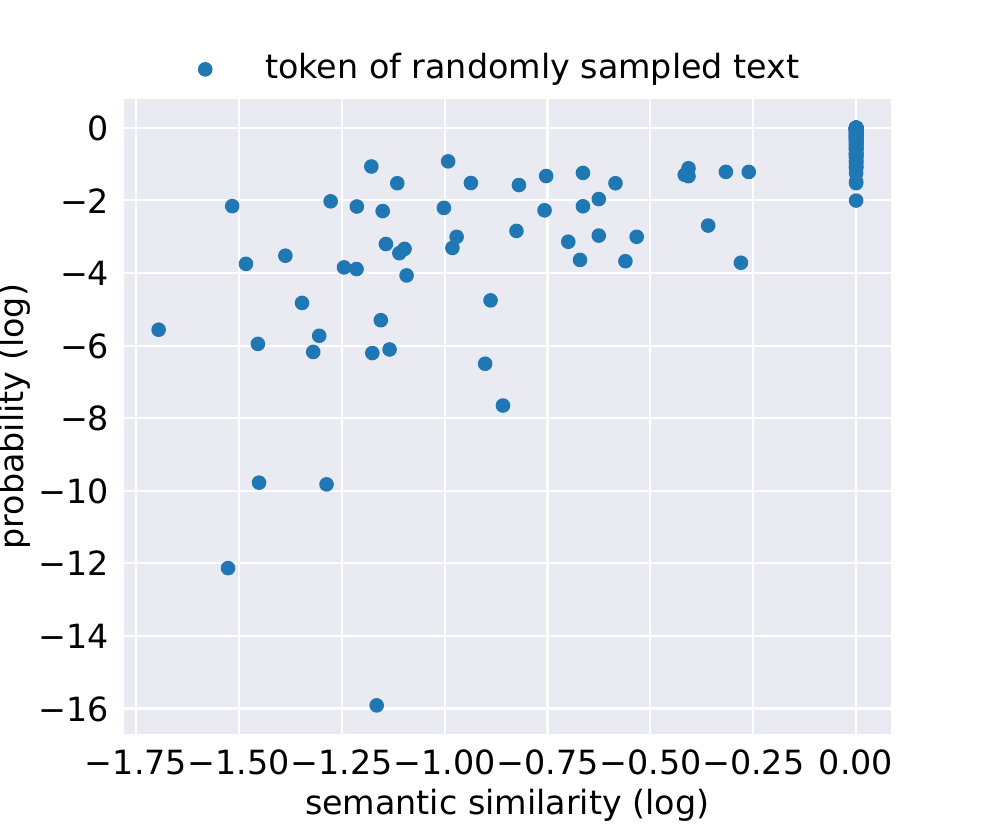} }
    \subfloat[Text: \textcolor{red}{``The 2014 Berlin…''}]{\includegraphics[width=1.7in]{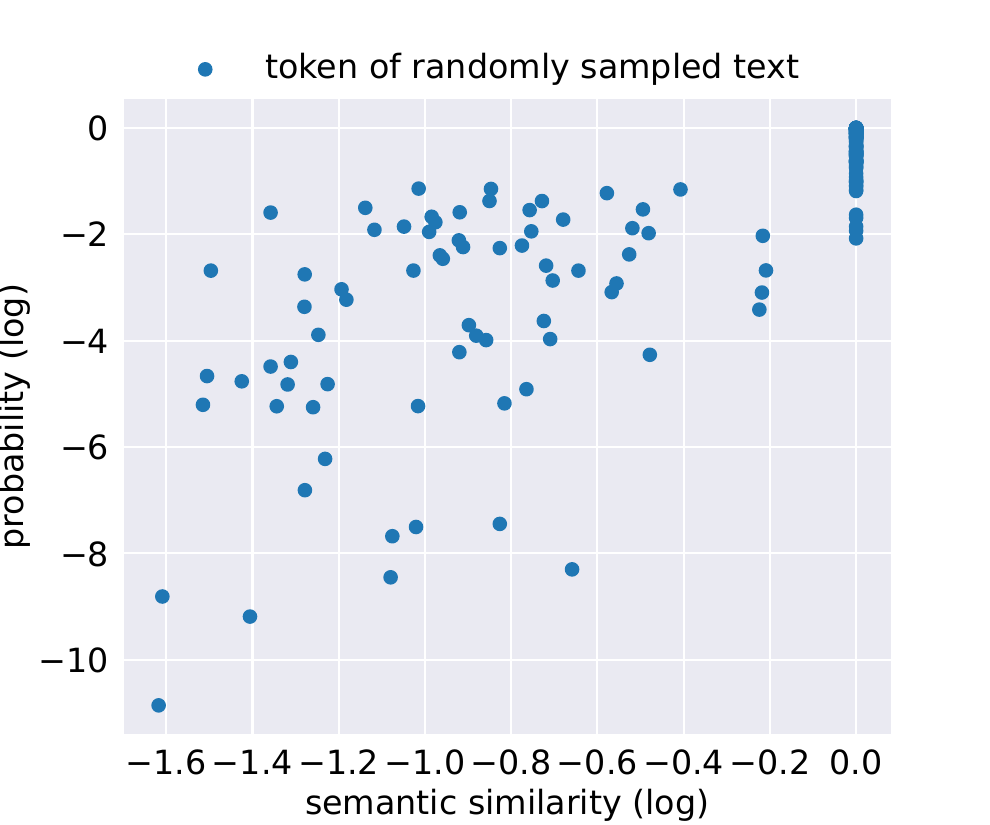} }
    \caption{The distribution of semantic similarity and probability of tokens from two randomly sampled text. }
    \label{Figure: pair distribution}
\vspace{-1em}
\end{figure}

\subsection{Approximating Output Probabilities via Token-level Semantic Similarity}
\label{sec: 4.2}
Given the sequence \(x=t_1,\ldots,t_n\), our strategy is to compute label-only ``proxies'' for probability of each token (i.e., \(f_\theta(t_i\mid t_1,\ldots,t_{i-1}), 1\leq i\leq n\)) by extracting information from the first generated token (i.e., \(\hat{t}_{i}\sim f_\theta(t_{i}\mid t_1,\ldots,t_{i-1}), 1\leq i\leq n\)). Following the perspective that language models tend to assign similar probabilities to semantically similar tokens \cite{bengio2000neural}, we naturally hypothesize that \(f_\theta(t_i\mid t_1,\ldots,t_{i-1})\) is potentially correlated with the semantic similarity between \(\hat{t}_{i}\) and \(t_{i}\) (i.e., \(sim(t_i, \hat{t}_i)\)). Moreover, as the model has assigned a high probability to \(\hat{t}_{i}\) (otherwise it would not have been sampled), higher \(sim(t_i, \hat{t}_i)\) implies higher likelihood of \(t_{i}\). We will use another toy example to illustrate this.

\partitle{Example 2} 
Assume we want to infer whether ``We all love Donald Duck.'' is a member. When feeding ``We all'' into the model, the next generated token being ``like'' instead of ``hate'' clearly suggests a higher \(f_\theta(\text{love}\mid \text{We all})\), even if both tokens are not the ground-truth.

To verify this, we collect the semantic similarity-probability pair (log scale) of each token for all samples from WikiMIA. Figure \ref{Figure: pair distribution} illustrates the distribution of semantic similarity and probability of tokens from two randomly sampled examples in WikiMIA, which presents a clear trend where probability increases with rising semantic similarity. We further calculate the average Pearson correlation coefficient between these two variables across all samples, and we get \textbf{0.765} with a p-value less than \textbf{0.001} on data from WikiMIA, indicating \textbf{the feasibility of using univariate linear regression for approximating output probabilities via semantic similarity.} 

The challenge is that the adversary cannot access the output logits on the target model, making it hard to obtain the regression results (i.e., regression parameters). To address this issue, we can conduct pre-regression on an open-source LLM (surrogate model) to acquire the required parameters. We summarize our method in Algorithm~\ref{alg:aop}. Our method's effectiveness stems from leveraging the specific properties of LLMs: Tokens (i.e., hard labels) generated by LLMs contain rich semantic information, which can be used to approximate output probabilities.

\begin{algorithm}[t]
  \microtypesetup{protrusion=false} 
    
  \SetKwInOut{KwIn}{Input}
  \SetKwInOut{KwOut}{Output}
  \KwIn{Sequence $x=t_1,\ldots,t_n$}
  \KwOut{Approximated probabilities Prob$(x)$ = $\log\tilde{p}(t_i\mid t_1,\ldots,t_{i-1})$ \\for \(i \in \{1, \ldots, n\}\)}
  \begin{spacing}{1}
  \For{\(i \in \{1, \ldots, n\}\)}
  {
   Compute $sim(t_i, \hat{t}_i)$ on Surrogate Model.\\
   Add $(sim_i, \log f(t_i\mid t_1,\ldots,t_{i-1}))$ \\to Surrogate$(x)$.
  }
  \end{spacing}
  Obtain $\beta$(slope) and $\alpha$(intercept) after regression on Surrogate$(x)$.
  \begin{spacing}{1}
  \For{\(i \in \{1, \ldots, n\}\)}
  {
   Compute $sim(t_i, \hat{t}_i)$ on Target Model.\\
   Add $\beta\cdot sim_i + \alpha$ to Prob$(x)$.
  }
  \end{spacing}
  return Prob$(x)$
  \caption{Approximating Probabilities}
  \label{alg:aop}

\end{algorithm}

\begin{table}[!t]
\newcommand{\tabincell}[2]{\begin{tabular}{@{}#1@{}}#2\end{tabular}}
\centering
\setlength{\tabcolsep}{2.5pt}
\caption{Regression results of semantic similarity-probability pairs on different pre-trained LLMs.}
\vspace{-2mm}
\scalebox{1.0}
{
\begin{tabular}{l|ccccc}
\toprule
Parameters &GPT-2 xl &Pythia-2.8B &phi-2 &gemma-2\\
\midrule
slope& 3.83& 3.68& 3.66& 3.47\\
intercept& -0.78& -0.65& -0.57& -0.60\\
\bottomrule
\end{tabular}
}
\label{tab: regression results}
\end{table}

\subsection{Membership Inference based on Approximated Perplexity}
\label{sec: 4.3}
To infer membership, the adversary can compute the approximated perplexity based on the Prob$(x)$ obtained from Section~\ref{sec: 4.2} as:
\begin{equation}
    \tilde{\mathcal{P}}=\exp(-\frac1{n}\sum_{i=1}^n\log\tilde{p}(t_i\mid t_1,\ldots,t_{i-1})).
\label{eq: perplexity}
\end{equation}

Since the model will assign a lower perplexity to the member sequence, we can detect them by thresholding this result. It is worth noting that although differences in the properties between the surrogate and target models can introduce errors in Prob$(x)$, experimental results show that this is already effective. We attribute this to the following potential reasons.

\partitle{Reason 1} 
We believe that the surrogate model mainly captures the mapping between semantic similarity and output probabilities, which has less relevance to the inherent properties of the model itself. To empirically verify this, we select four open-source pre-trained LLMs which differ significantly in pre-training datasets, architectures, and release times: GPT-2 xl \cite{radford2019language}, Pythia-2.8B \cite{biderman2023pythia}, phi-2 \cite{textbooks2} and gemma-2-2B \cite{team2024gemma}. Using data from WikiMIA, we calculate the regression results (i.e., the slope and the intercept) of semantic similarity-probability pairs on these models. Table \ref{tab: regression results} indicates that the mapping between semantic similarity and output probabilities is indeed close across different models.

\partitle{Reason 2} 
We believe that token-level semantic similarity indeed serves as a good substitute for output probabilities, whereas the main role of the surrogate model is to \textbf{align the distribution of semantic similarity scores with the true output probabilities.} Specifically, due to the calculation principles of the semantic similarity metric, its range is much narrower than that of output probabilities, as shown in Figure \ref{Figure: pair distribution}. This makes low-probability tokens have higher membership scores due to the high lower bound of semantic similarity scores, which might negatively impact performance compared to using probabilities. Even with a slight gap between the regression results on the surrogate model and the target model, the surrogate model can already align the two distributions.

\noindent We leave the empirical evaluation on the impact of different surrogate models in Section~\ref{sec: 5.3}. 

\vspace{-0.5em}
\section{Experiments}

In this section, we evaluate PETAL on two benchmark datasets and diverse pre-trained LLMs. We focus on three standard metrics: Balanced Accuracy, AUC, and TPR at low FPR, which have been detailed in Section \ref{sec: 3.1}. Through extensive experiments, we demonstrate that our attack outperforms existing label-only attacks and performs on par with advanced logits-based attacks. Following that, the correlation between PETAL and its modifiable parameters is presented and discussed. Specifically, we show how PETAL performs when the value of a particular parameter is changed. We finally conduct two case studies on fine-tuned LLMs to demonstrate the generalizability of PETAL in fine-tuning scenarios.

\subsection{Attack Setup}

\partitle{Datasets} We focus on two datasets commonly used in the task of MIAs against pre-trained LLMs: WikiMIA \cite{shi2023detecting} and MIMIR \cite{duan2024membership}. WikiMIA is the first dataset designed for evaluating pre-training data detection, consisting of events from Wikipedia. According to the event dates and the target model's release date, texts can be automatically classified as members or non-members. MIMIR is built upon the Pile dataset \cite{gao2020pile} by Duan et al. \cite{duan2024membership}, where training samples and non-training samples are drawn from its training and test split, respectively. Note that Duan et al. argue that WikiMIA has distribution shifts and temporal discrepancies between members and non-members, which differs from the classical MIA game \cite{yeom2018privacy}. However, we believe that distinguishing between members and temporally shifted non-members has practical implications in the context of MIAs against pre-trained LLMs. Therefore, following previous work \cite{shi2023detecting,zhang2024min,xie2024recall}, we report results on both benchmarks to ensure experimental integrity.

\partitle{Models} WikiMIA is suitable for a wide range of pre-trained LLMs because data from Wikipedia is typically included in existing models' pre-training corpus. For evaluations on WikiMIA, we select LLMs released after 2021, such as LLaMA2-13B, Falcon-7B, Pythia-6.9B, and OPT-6.7B, due to their publicly available time cutoffs of pre-training data. As MIMIR is built upon the Pile, it is only applicable to models trained on the original Pile data. Following Duan et al. \cite{duan2024membership}, we focus on the Pythia family (160M, 1.4B, 2.8B, 6.9B). When a reference model is needed to evaluate certain attacks, we do not choose smaller versions of the target model \cite{shi2023detecting} (e.g., using Pythia-160M for Pythia-6.9B). Instead, we consistently use GPT-2 xl \cite{radford2019language} as the reference model, since smaller versions of the target model may not exist or may be inaccessible to the adversary in real scenarios.

\begin{table*}[!t]
\centering
\setlength{\tabcolsep}{3pt}
\caption{Complete results of various attacks on WikiMIA \cite{shi2023detecting}. We bold our results when they surpass all label-only attacks and underline our results when they are comparable to logits-based attacks (i.e., surpassing at least two logits-based attacks).}
\vspace{-2mm}
\scalebox{0.95}
{
\begin{tabular}{l|cccccccccccc}
\toprule  
Metrics & \multicolumn{4}{c}{AUC $\uparrow$}&  \multicolumn{4}{c}{Balanced Acc $\uparrow$}& \multicolumn{4}{c}{TPR@1\%FPR  $\uparrow$}\\
\cmidrule(l{5pt}r{5pt}){2-5}\cmidrule(l{5pt}r{5pt}){6-9}\cmidrule(l{5pt}r{5pt}){10-13}
Models & Pythia& OPT& LLaMA2& Falcon&Pythia& OPT& LLaMA2& Falcon& Pythia& OPT& LLaMA2& Falcon\\
\midrule
\textbf{\textit{Logits-based Attacks}}&&&&&&&&&&&&\\
PPL attack              &0.64&0.61&0.58&0.60&0.62&0.58&0.56&0.58&6.2\%&3.4\%&1.3\%&1.0\%\\
reference attack        &0.50&0.45&0.40&0.43&0.52&0.50&0.50&0.51&2.1\%&0.3\%&0.0\%&0.0\%\\
zlib attack             &0.64&0.61&0.56&0.60&0.62&0.59&0.56&0.58&4.9\%&3.4\%&2.3\%&3.4\%\\
neighborhood attack     &0.66&0.64&0.58&0.62&0.62&0.62&0.57&0.60&0.8\%&0.8\%&0.6\%&2.6\%\\
MIN-K\% PROB            &0.66&0.62&0.53&0.57&0.63&0.61&0.53&0.56&8.8\%&4.4\%&0.8\%&4.1\%\\
\midrule
\textbf{\textit{Label-only Attacks}}&&&&&&&&&&&&\\
WS attack               &0.49&0.47&0.51&0.51&0.52&0.51&0.54&0.52&0.3\%&0.3\%&0.3\%&0.8\%\\
RS attack               &0.50&0.47&0.50&0.51&0.51&0.50&0.52&0.52&0.3\%&0.5\%&0.5\%&0.3\%\\
BT attack               &0.51&0.48&0.49&0.48&0.53&0.50&0.52&0.51&1.0\%&0.5\%&1.6\%&0.5\%\\
PETAL (Ours)&\textbf{\ul{0.64}}&\textbf{\ul{0.62}}&\textbf{\ul{0.58}}&\textbf{\ul{0.60}}&\textbf{\ul{0.62}}&\textbf{\ul{0.59}}&\textbf{\ul{0.57}}&\textbf{\ul{0.57}}&\textbf{\ul{4.9\%}}&\textbf{\ul{3.1\%}}&\textbf{\ul{1.6\%}}&\textbf{\ul{2.1\%}}\\
\bottomrule
\end{tabular}
}
\vspace{-0.1em}
\label{table:main results on WikiMIA}
\end{table*}

\begin{table*}[!t]
\centering
\setlength{\tabcolsep}{2pt}
\caption{AUC results of various attacks on different subsets of the Pile \cite{gao2020pile}.}
\vspace{-2mm}
\scalebox{0.97}
{
\begin{tabular}{l|cccccccccccccccc}
\toprule
Subsets & \multicolumn{4}{c}{arXiv}&  \multicolumn{4}{c}{DM Mathematics}& \multicolumn{4}{c}{GitHub}& \multicolumn{4}{c}{HackerNews}\\
\cmidrule(l{5pt}r{5pt}){2-5}\cmidrule(l{5pt}r{5pt}){6-9}\cmidrule(l{5pt}r{5pt}){10-13}\cmidrule(l{5pt}r{5pt}){14-17}
Models & 160M& 1.4B& 2.8B& 6.9B& 160M& 1.4B& 2.8B& 6.9B& 160M& 1.4B& 2.8B& 6.9B& 160M& 1.4B& 2.8B& 6.9B\\
\midrule
\textbf{\textit{Logits-based Attacks}}&&&&&&&&&&&&\\
PPL attack&0.60&0.66&0.64&0.66&0.91&0.88&0.87&0.86&0.84&0.86&0.87&0.88&0.59&0.59&0.60&0.60\\
reference attack&0.63&0.67&0.64&0.67&0.83&0.82&0.78&0.77&0.65&0.65&0.65&0.64&0.53&0.52&0.53&0.52\\
zlib attack&0.62&0.65&0.64&0.66&0.82&0.80&0.78&0.77&0.83&0.85&0.86&0.87&0.59&0.59&0.60&0.59\\
neighborhood attack&0.67&0.69&0.69&0.69&0.70&0.73&0.72&0.73&0.80&0.83&0.83&0.84&0.56&0.53&0.55&0.54\\
MIN-K\% PROB&0.59&0.62&0.61&0.62&0.71&0.71&0.67&0.68&0.83&0.86&0.86&0.88&0.57&0.57&0.58&0.59\\
\textbf{\textit{Label-only Attacks}}&&&&&&&&&&&&\\
WS attack&0.47&0.50&0.50&0.50&0.50&0.45&0.49&0.44&0.71&0.76&0.76&0.77&0.49&0.51&0.51&0.48\\
RS attack&0.48&0.50&0.48&0.51&0.40&0.42&0.45&0.43&0.74&0.77&0.77&0.78&0.48&0.49&0.51&0.51\\
BT attack&0.53&0.51&0.51&0.55&0.40&0.39&0.45&0.39&0.71&0.74&0.76&0.77&0.50&0.51&0.52&0.53\\
PETAL (Ours)&\textbf{0.53}&\textbf{0.58}&\textbf{0.57}&\textbf{0.58}&\ul{\textbf{0.86}}&\ul{\textbf{0.87}}&\ul{\textbf{0.85}}&\ul{\textbf{0.85}}&\ul{\textbf{0.83}}&\ul{\textbf{0.85}}&\ul{\textbf{0.85}}&\ul{\textbf{0.87}}&\ul{\textbf{0.59}}&\ul{\textbf{0.59}}&\ul{\textbf{0.58}}&\ul{\textbf{0.58}}\\
\vspace{-.9em}\\
\toprule
Subsets & \multicolumn{4}{c}{Pile CC}&  \multicolumn{4}{c}{PubMed Central}& \multicolumn{4}{c}{Wikipedia}& \multicolumn{4}{c}{Average}\\
\cmidrule(l{5pt}r{5pt}){2-5}\cmidrule(l{5pt}r{5pt}){6-9}\cmidrule(l{5pt}r{5pt}){10-13}\cmidrule(l{5pt}r{5pt}){14-17}
Models & 160M& 1.4B& 2.8B& 6.9B& 160M& 1.4B& 2.8B& 6.9B& 160M& 1.4B& 2.8B& 6.9B& 160M& 1.4B& 2.8B& 6.9B\\
\midrule
\textbf{\textit{Logits-based Attacks}}&&&&&&&&&&&&\\
PPL attack&0.51&0.52&0.53&0.54&0.71&0.69&0.68&0.70&0.60&0.62&0.63&0.64&0.68&0.69&0.69&0.70\\
reference attack&0.49&0.51&0.51&0.53&0.67&0.65&0.64&0.65&0.54&0.58&0.59&0.59&0.62&0.63&0.62&0.62\\
zlib attack&0.50&0.51&0.52&0.52&0.73&0.71&0.70&0.72&0.56&0.60&0.61&0.62&0.66&0.67&0.67&0.68\\
neighborhood attack&0.49&0.50&0.50&0.51&0.70&0.69&0.68&0.69&0.53&0.56&0.56&0.57&0.64&0.65&0.65&0.65\\
MIN-K\% PROB&0.51&0.52&0.52&0.52&0.58&0.53&0.53&0.55&0.57&0.59&0.59&0.62&0.62&0.63&0.62&0.64\\
\textbf{\textit{Label-only Attacks}}&&&&&&&&&&&&\\
WS attack&0.50&0.50&0.49&0.50&0.51&0.52&0.54&0.54&0.50&0.53&0.55&0.52&0.53&0.54&0.55&0.54\\
RS attack&0.50&0.50&0.48&0.49&0.53&0.54&0.54&0.56&0.51&0.54&0.54&0.55&0.52&0.54&0.54&0.55\\
BT attack&0.48&0.52&0.50&0.50&0.53&0.52&0.53&0.56&0.52&0.53&0.54&0.52&0.52&0.53&0.54&0.55\\
PETAL (Ours)&\ul{\textbf{0.53}}&\ul{\textbf{0.54}}&\ul{\textbf{0.55}}&\ul{\textbf{0.54}}&\textbf{0.64}&\ul{\textbf{0.65}}&\ul{\textbf{0.65}}&\ul{\textbf{0.66}}&\ul{\textbf{0.58}}&\ul{\textbf{0.62}}&\ul{\textbf{0.61}}&\ul{\textbf{0.61}}&\ul{\textbf{0.65}}&\ul{\textbf{0.67}}&\ul{\textbf{0.67}}&\ul{\textbf{0.67}}\\
\bottomrule
\end{tabular}
}
\vspace{-1.5em}
\label{table:main results auc on MIMIR}
\end{table*}

\begin{table*}[!t]
\centering
\setlength{\tabcolsep}{3.6pt}
\caption{TPR results of various attacks on different subsets of the Pile \cite{gao2020pile}.}
\vspace{-2mm}
\scalebox{0.78}
{
\begin{tabular}{l|cccccccccccccccc}
\toprule
Subsets & \multicolumn{4}{c}{arXiv}&  \multicolumn{4}{c}{DM Mathematics}& \multicolumn{4}{c}{GitHub}& \multicolumn{4}{c}{HackerNews}\\
\cmidrule(l{5pt}r{5pt}){2-5}\cmidrule(l{5pt}r{5pt}){6-9}\cmidrule(l{5pt}r{5pt}){10-13}\cmidrule(l{5pt}r{5pt}){14-17}
Models & 160M& 1.4B& 2.8B& 6.9B& 160M& 1.4B& 2.8B& 6.9B& 160M& 1.4B& 2.8B& 6.9B& 160M& 1.4B& 2.8B& 6.9B\\
\midrule
\textbf{\textit{Logits-based Attacks}}&&&&&&&&&&&&\\
PPL attack&5.6\%&10.4\%&6.8\%&10.4\%&57.3\%&51.7\%&34.8\%&42.7\%&30.4\%&35.6\%&43.2\%&48.8\%&5.2\%&4.8\%&6.0\%&5.6\%\\
reference attack&1.6\%&2.4\%&2.0\%&1.6\%&57.3\%&49.4\%&50.6\%&47.2\%&8.8\%&15.2\%&12.8\%&9.6\%&2.4\%&1.2\%&1.2\%&1.2\%\\
zlib attack&9.2\%&10.8\%&10.8\%&12.8\%&50.6\%&41.6\%&23.6\%&38.2\%&17.2\%&27.6\%&31.2\%&36.8\%&4.8\%&4.8\%&4.4\%&5.2\%\\
neighborhood attack&4.8\%&14.8\%&12.4\%&12.0\%&5.6\%&2.2\%&1.1\%&4.5\%&22.4\%&32.0\%&26.0\%&27.2\%&2.4\%&0.8\%&2.4\%&1.2\%\\
MIN-K\% PROB&3.2\%&4.4\%&1.6\%&4.4\%&24.7\%&21.3\%&19.1\%&12.4\%&26.4\%&23.6\%&31.6\%&28.4\%&2.0\%&3.2\%&4.8\%&4.8\%\\
\textbf{\textit{Label-only Attacks}}&&&&&&&&&&&&\\
WS attack&1.2\%&1.6\%&1.2\%&0.0\%&0.0\%&0.0\%&2.2\%&1.1\%&10.4\%&13.6\%&17.2\%&21.6\%&0.8\%&0.4\%&1.2\%&2.0\%\\
RS attack&1.2\%&1.2\%&0.8\%&0.8\%&0.0\%&0.0\%&2.2\%&0.0\%&8.4\%&28.8\%&32.0\%&35.6\%&0.0\%&0.4\%&0.8\%&0.4\%\\
BT attack&0.8\%&0.4\%&2.8\%&1.6\%&0.0\%&1.1\%&3.4\%&3.4\%&10.4\%&22.8\%&25.2\%&25.2\%&0.4\%&0.0\%&0.4\%&0.0\%\\
PETAL (Ours)&\textbf{2.4\%}&\ul{\textbf{5.2\%}}&\ul{\textbf{4.8\%}}&\ul{\textbf{6.4\%}}&\ul{\textbf{44.9\%}}&\ul{\textbf{33.7\%}}&\ul{\textbf{37.1\%}}&\ul{\textbf{29.2\%}}&\ul{\textbf{26.4\%}}&\ul{\textbf{31.2\%}}&\ul{\textbf{33.6\%}}&\ul{\textbf{44.4\%}}&\ul{\textbf{8.4\%}}&\ul{\textbf{3.6\%}}&\ul{\textbf{4.4\%}}&\ul{\textbf{6.8\%}}\\
\vspace{-.9em}\\
\toprule
Subsets & \multicolumn{4}{c}{Pile CC}&  \multicolumn{4}{c}{PubMed Central}& \multicolumn{4}{c}{Wikipedia}& \multicolumn{4}{c}{Average}\\
\cmidrule(l{5pt}r{5pt}){2-5}\cmidrule(l{5pt}r{5pt}){6-9}\cmidrule(l{5pt}r{5pt}){10-13}\cmidrule(l{5pt}r{5pt}){14-17}
Models & 160M& 1.4B& 2.8B& 6.9B& 160M& 1.4B& 2.8B& 6.9B& 160M& 1.4B& 2.8B& 6.9B& 160M& 1.4B& 2.8B& 6.9B\\
\midrule
\textbf{\textit{Logits-based Attacks}}&&&&&&&&&&&&\\
PPL attack&3.2\%&4.0\%&4.4\%&5.6\%&9.6\%&4.8\%&4.8\%&2.0\%&1.6\%&4.8\%&4.0\%&3.2\%&16.1\%&16.6\%&14.9\%&16.9\%\\
reference attack&0.8\%&3.2\%&2.8\%&3.2\%&4.4\%&2.0\%&1.2\%&1.6\%&3.2\%&3.6\%&0.8\%&1.6\%&11.2\%&11.0\%&10.2\%&9.4\%\\
zlib attack&2.0\%&4.4\%&5.6\%&6.4\%&10.0\%&5.2\%&3.6\%&3.2\%&2.4\%&2.0\%&2.4\%&2.4\%&13.7\%&13.8\%&11.7\%&15.0\%\\
neighborhood attack&2.8\%&3.2\%&3.2\%&5.2\%&13.2\%&8.4\%&7.2\%&5.2\%&0.4\%&0.4\%&0.8\%&1.2\%&7.4\%&8.8\%&7.6\%&8.1\%\\
MIN-K\% PROB&1.2\%&3.6\%&2.8\%&4.4\%&2.0\%&1.6\%&1.2\%&1.6\%&3.2\%&2.0\%&0.4\%&2.0\%&9.0\%&8.5\%&8.8\%&8.3\%\\
\textbf{\textit{Label-only Attacks}}&&&&&&&&&&&&\\
WS attack&3.6\%&1.2\%&0.4\%&2.4\%&3.2\%&0.4\%&1.2\%&0.8\%&2.0\%&2.8\%&1.6\%&4.4\%&3.0\%&2.9\%&3.6\%&4.6\%\\
RS attack&2.8\%&0.0\%&2.4\%&0.4\%&3.2\%&0.8\%&2.0\%&2.8\%&2.4\%&3.2\%&2.0\%&3.6\%&2.6\%&4.9\%&6.0\%&6.2\%\\
BT attack&0.8\%&2.4\%&2.4\%&2.0\%&2.8\%&3.6\%&4.4\%&4.0\%&2.0\%&6.0\%&2.8\%&3.6\%&2.5\%&5.2\%&5.9\%&5.7\%\\
PETAL (Ours)&\ul{\textbf{3.6\%}}&\ul{\textbf{4.0\%}}&\ul{\textbf{4.4\%}}&\ul{\textbf{5.6\%}}&\ul{\textbf{4.8\%}}&\ul{\textbf{6.0\%}}&\ul{4.0\%}&\ul{\textbf{4.4\%}}&\ul{1.6\%}&\ul{3.2\%}&\ul{\textbf{2.8\%}}&\ul{4.0\%}&\ul{\textbf{13.1\%}}&\ul{\textbf{12.4\%}}&\ul{\textbf{13.0\%}}&\ul{\textbf{14.4\%}}\\
\bottomrule
\end{tabular}
}
\label{table:main results TPR on MIMIR}
\end{table*}

\partitle{Baselines} We compare our PETAL with six state-of-the-art or representative attack methods. For logits-based baselines, we select PPL attack \cite{carlini2021extracting,yeom2018privacy}, reference attack \cite{carlini2021extracting}, zlib attack \cite{carlini2021extracting}, neighborhood attack \cite{mattern-etal-2023-membership}, and MIN-K\% PROB \cite{shi2023detecting}. For label-only baselines, we mainly consider the three variants of the attacks by Tang et al. \cite{tang2023assessing}, which evaluate robustness using data augmentation. We do not consider attacks that evaluate robustness by computing the adversarial perturbations \cite{choquette2021label,li2021membership,wu2024you}, since they cannot be directly applied to generative LLMs. Specifically, these attacks compute the minimum perturbation required to change the model's prediction and identify members based on the findings that adversarial perturbations tend to be larger for members than for non-members. However, in generative LLMs, even a single token change in the input can alter the next token prediction, whether for members or non-members. Therefore, adversarial perturbations cannot serve as effective MIA signals.

All of the aforementioned attacks have been detailed in Section \ref{sec: 3.2} \& \ref{sec: 3.3}. Note that some reference-based attacks consider the difficulty of the samples and could potentially achieve better performance. In the main experimental section, we do not apply these techniques to PETAL (which can be directly applied actually), as our primary goal is to show that output probabilities are not necessary for MIAs against LLMs.

\partitle{Other Implementation Details} We obtain all pre-trained LLMs using the Huggingface transformers library \cite{wolf2020transformers} and PyTorch \cite{paszke2019pytorch}. In the main experiments, we uniformly truncate samples to a length of 32 words to avoid the influence of text length on attack performances. For evaluating the neighborhood attack, we use 10 neighbor samples for every text input to calculate its intrinsic hardness. As synthetically generating perturbed versions of text input can be time-consuming, we use perturbed WikiMIA\footnote{\url{https://huggingface.co/datasets/zjysteven/WikiMIA_paraphrased_perturbed}} by Zhang et al. \cite{zhang2024min} and perturbed MIMIR\footnote{\url{https://huggingface.co/datasets/iamgroot42/mimir}} by Duan et al. \cite{duan2024membership}. For evaluating MIN-K\% PROB, we set $k=20$ as done in the original paper. For evaluating robustness-based attacks, we sweep over the range of 10\% to 90\% for $n\%$ to report the optimal result. Since robustness-based attacks require training shadow models \cite{shokri2017membership} to generate training data for training a classifier, which is unrealistic for pre-trained LLMs, we instead compute the average text similarity scores and attack by thresholding \cite{yeom2018privacy}. When evaluating our PETAL, we adopt a sentence transformer to capture semantic similarity. Specifically, we use all-MiniLM-L6-v2 \cite{lewis2020bart} to map tokens to normalized vectors and compute the dot product between two vectors as the token-level semantic similarity. We utilize GPT-2 xl \cite{radford2019language} as the surrogate model because 1) GPT-2 is an early model trained on WebText rather than the Pile, and 2) WebText excludes data from Wikipedia. \textbf{This sufficiently eliminates concerns that PETAL benefits from potential member privacy leakage of the surrogate model}. In our main experiments, the decoding strategy is set as greedy search. All these factors that could potentially impact the attack results will be extensively discussed in Section~\ref{sec: 5.3}. To evaluate our attack's TPR for predetermined low FPR (1\%), we adjust the threshold $\tau$ to meet the requirement. For stochastic experiments (not using deterministic decoding strategies), we run the experiments 3 times and report the average results. All models are deployed on 8 GeForce RTX 3090 GPUs. When the model is too large to fit into the GPU memory, we use the half-precision floating-point version.

\subsection{Main Results}
Here we show the detailed attack results with a comparison to five advanced logits-based attacks and three label-only attacks. Complete results on WikiMIA are summarized in Table \ref{table:main results on WikiMIA}. Due to page limits, AUC results on MIMIR are summarized in Table \ref{table:main results auc on MIMIR}, and TPR results on MIMIR are summarized in Table \ref{table:main results TPR on MIMIR} respectively. For balanced accuracy results on MIMIR, please see Appendix \ref{sec: main Acc results}. For complete ROC curves of all attacks on MIMIR and WikiMIA, please see Appendix \ref{sec: roc curves}. We believe our extensive empirical evaluation introduces some insightful findings into MIAs against pre-trained LLMs, which we discuss in detail below.

\partitle{Robustness-based Attacks Are Nearly Random}
As we have claimed in Section \ref{sec: 3.3}, robustness estimated from perturbations likely cannot serve as a reliable indicator for inferring membership, because token-level perturbations are too coarse to capture the minor gap in distance between members and non-members to the decision boundary. The results in Table \ref{table:main results on WikiMIA} and Table \ref{table:main results auc on MIMIR} confirm this again—all three robustness-based attacks have achieved an AUC close to 0.5 across models of different sizes, with the exception of GitHub (even on this subset, robustness-based attacks are much weaker than logits-based attacks). We attribute the reason robustness-based attacks are effective on the GitHub subset to the higher sample difficulty of code data. Thus, LLMs tend to be particularly uncertain with unseen code data, increasing the robustness gap between members and non-members, making token-level perturbations fine-grained enough to capture it.

\begin{table}[!t]
\newcommand{\tabincell}[2]{\begin{tabular}{@{}#1@{}}#2\end{tabular}}
\centering
\setlength{\tabcolsep}{2.5pt}
\caption{Performance comparison of PETAL under different query settings on WikiMIA
\cite{shi2023detecting}.}
\vspace{-1mm}
\scalebox{1.0}
{
\begin{tabular}{c|ccccc}
\toprule
Query ($\frac{m}{n}$)&AUC$\uparrow$ &Balanced Acc$\uparrow$&TPR@1\%FPR$\uparrow$ \\
\midrule
25\%& 0.57& 0.57& 2.1\%\\
50\%& 0.61& 0.59& 3.1\%\\
75\%& 0.62& 0.59& 3.1\%\\
100\%& 0.64& 0.62& 4.9\%\\
\bottomrule
\end{tabular}
}
\vspace{-0.5em}
\label{table: query budget on WikiMIA}
\end{table}

\partitle{PETAL Are Comparable to Logits-based Attacks}
Tables \ref{table:main results on WikiMIA} and \ref{table:main results auc on MIMIR} show that our PETAL significantly outperforms other label-only attacks across all datasets and target model settings. Regarding comparison with existing logits-based attacks, since it's hard to select an overall best attack, we consider PETAL comparable to logits-based attacks when it surpasses at least two other methods. It can be observed that PETAL \textbf{always} performs on par with logits-based attacks, solely utilizing hard labels. The only exception is on the arXiv subset---despite being superior to existing label-only attacks, PETAL lags behind all logits-based attacks. One possible reason is that the arXiv subset contains extensive LaTeX source codes and special characters, making it challenging for GPT2-xl to map semantic similarity to probabilities accurately. In other words, GPT2-xl may struggle to generate meaningful tokens to approximate output probabilities when prompted with obscure prefixes (like LaTeX codes). From another perspective, the performance differences between PETAL and the PPL attack, which uses output probabilities directly, are relatively marginal (AUC differences within ±0.03), highlighting PETAL's effectiveness as a label-only attack.

\partitle{Breakthrough in TPR at Low FPR}
Remarkably, Carlini et al. \cite{carlini2022membership} have raised doubts about MIAs that rely on limited information (e.g., hard labels), suggesting they might not achieve high success rates at low FPR. Table \ref{table:main results TPR on MIMIR} indicates that these doubts are somewhat justified, as robustness-based attacks indeed show significantly lower TPR at 1\% FPR compared to logits-based attacks on some datasets (e.g., the PPL attack gets at least 15x TPR at 1\% FPR than that of robustness-based attacks on the DM Mathematics subset). However, our PETAL has achieved a breakthrough in TPR at low FPR without access to output probabilities. Specifically, Table \ref{table:main results TPR on MIMIR} demonstrates that PETAL achieves comparable TPR at 1\% FPR to logits-based attacks across all datasets, including the DM Mathematics subset where all robustness-based attacks fail. Our findings indicate that even in the label-only setting, pre-trained LLMs face an equal level of privacy threat. 

\partitle{Larger Models Memorize Better}
Table \ref{table:main results auc on MIMIR} shows a general trend---the performance of MIAs slightly improves as the model size increases, which is consistent with previous findings \cite{carlini2021extracting}. This can be attributed to the faster and better memorization capabilities of larger pre-trained LLMs~\cite{tirumala2022memorization}. Nonetheless, our PETAL consistently performs on par with other logits-based attacks, demonstrating its robustness to the varying memorization capabilities of LLMs.

\subsection{Parameter Analysis}
\label{sec: 5.3}

In this section, we conduct extensive experiments to investigate the specific impact of each parameter on the final performance. We aim to 1) further substantiate our explanation in Section \ref{sec: 4} regarding the effectiveness of our PETAL. 2) demonstrate the robustness of PETAL. Specifically, we start by exploring the impact of query budget. Then we discuss the effects of the granularity of probability estimation. We also carefully examine the impact of the selection of a surrogate model. Finally, we test PETAL under different text length, decoding strategy, weighting strategy and sentence transformer settings. In our parameter analysis experiments, we employ the Pythia-6.9B by default unless otherwise stated, since both WikiMIA and MIMIR can be used to evaluate attack performance on it.

\partitle{Different Query Budget}
We first explore how the query budget affects the performance of our PETAL. Recall that implementing our attack on a target sample with a length of $n$ tokens requires querying the target model $n$ times (see Algorithm \ref{alg:aop}). We reduce the query budget to $m$ by approximating the output probabilities of only the last $m$ tokens in the sample. Table \ref{table: query budget on WikiMIA} and Table \ref{table: query budget on MIMIR} show the attack results for different $\frac{m}{n}$ settings, including 1 (original PETAL), 0.75, 0.5, and 0.25. While a decrease in the number of queries generally leads to a decline in attack performance (with the exception of TPR at 1\% FPR), PETAL consistently outperforms robustness-based attacks. Moreover, with $\frac{m}{n} \geq 0.5$, our attack begins to perform on par with other logits-based attacks (see Table \ref{table:main results on WikiMIA} and Table \ref{table:main results auc on MIMIR}). We believe the reason TPR does not decrease with fewer queries is that this metric is more related to outlier tokens rather than the average probability over all tokens. Therefore, reducing the proportion of tokens for which probabilities need to be estimated does not necessarily lower the TPR.

\begin{table}[!t]
\newcommand{\tabincell}[2]{\begin{tabular}{@{}#1@{}}#2\end{tabular}}
\centering
\setlength{\tabcolsep}{2.5pt}
\caption{Performance comparison of PETAL under different query settings on MIMIR \cite{duan2024membership}.}
\vspace{-1mm}
\scalebox{1.0}
{
\begin{tabular}{c|ccccc}
\toprule
Query ($\frac{m}{n}$)&AUC$\uparrow$ &Balanced Acc$\uparrow$&TPR@1\%FPR$\uparrow$ \\
\midrule
25\%& 0.63& 0.62& 10.0\%\\
50\%& 0.65& 0.63& 12.2\%\\
75\%& 0.67& 0.65& 11.5\%\\
100\%& 0.67& 0.65& 14.4\%\\
\bottomrule
\end{tabular}
}
\vspace{-1em}
\label{table: query budget on MIMIR}
\end{table}

\begin{figure}[t]
    \centering
    \includegraphics[width=0.44\textwidth]{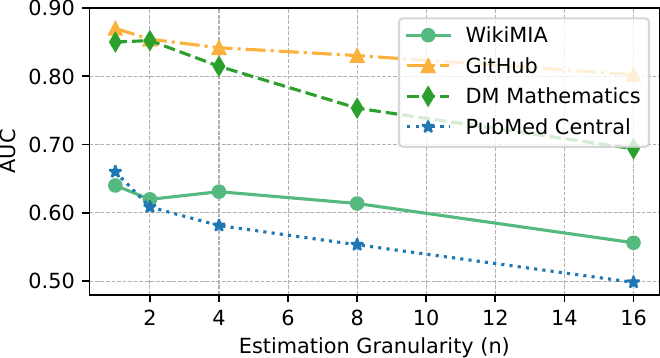}
    \caption{AUC values of our PETAL against Pythia-6.9B. The granularity of probability estimation is set as 1, 2, 4, 8, and 16. To avoid overcrowding the figure, the attack results on other subsets of MIMIR \cite{duan2024membership} can be found in Appendix \ref{sec: estimation granularity}.}
    \label{Figure: ablation on estimation granularity}
\vspace{-0.5em}
\end{figure}

\partitle{Different Granularity of Probability Estimation}
As we have claimed in Section \ref{sec: 4.1}, in a single query only the first generated token contains effective MIA signals. Our PETAL is built upon this hypothesis and has achieved empirical success in the main experiments. To further validate this claim, we design experiments to explore the impact of the granularity of probability estimation on attack results. Specifically, we simultaneously use the first \textbf{$n$} generated tokens to calculate semantic similarity and estimate output probabilities in a single query (i.e., using \(sim(t_{i+1}\oplus\vcenter{\hbox{\ldots}}\oplus t_{i+n}, \hat{t}_{i+1}\oplus\vcenter{\hbox{\ldots}}\oplus\hat{t}_{i+n})\) to estimate \(p(t_{i+1},\ldots,t_{i+n}\mid t_1,\ldots,t_i)\)). Using the chain rule of probability, we can obtain a coarser approximation of perplexity than that of Equation \ref{eq: perplexity} and attack by thresholding it. Figure \ref{Figure: ablation on estimation granularity} shows that with $n$ increasing from 1 to 16, PETAL's attack performance on WikiMIA and MIMIR gradually decreases. This is because the subsequent $n-1$ tokens in a single query contain little membership signal, which could have been extracted by generating only one token.

\begin{table*}[!t]
\centering
\setlength{\tabcolsep}{4.0pt}
\caption{Attack results of PETAL on WikiMIA \cite{shi2023detecting} under different decoding strategies.}
\vspace{-1mm}
\scalebox{0.95}
{
\begin{tabular}{l|cccccccccccc}
\toprule
Metrics & \multicolumn{4}{c}{AUC $\uparrow$}&  \multicolumn{4}{c}{Balanced Acc $\uparrow$}& \multicolumn{4}{c}{TPR@1\%FPR  $\uparrow$}\\
\cmidrule(l{5pt}r{5pt}){2-5}\cmidrule(l{5pt}r{5pt}){6-9}\cmidrule(l{5pt}r{5pt}){10-13}
Decoding Strategy & Pythia& OPT& LLaMA2& Falcon&Pythia& OPT& LLaMA2& Falcon& Pythia& OPT& LLaMA2& Falcon\\
\midrule
Greedy Search       &0.64&0.62&0.58&0.60&0.62&0.59&0.57&0.57&4.9\%&3.1\%&1.6\%&2.1\%\\
Nucleus Sampling    &0.61&0.61&0.58&0.60&0.59&0.58&0.56&0.58&1.6\%&4.9\%&1.0\%&2.6\%\\
Contrastive Search  &0.62&0.62&0.57&0.60&0.59&0.60&0.56&0.57&5.2\%&5.4\%&1.8\%&3.4\%\\
\bottomrule
\end{tabular}
}
\vspace{-1em}
\label{table: as_decoding}
\end{table*}

\begin{figure}[t]
    \centering
    \includegraphics[width=0.3\textwidth]{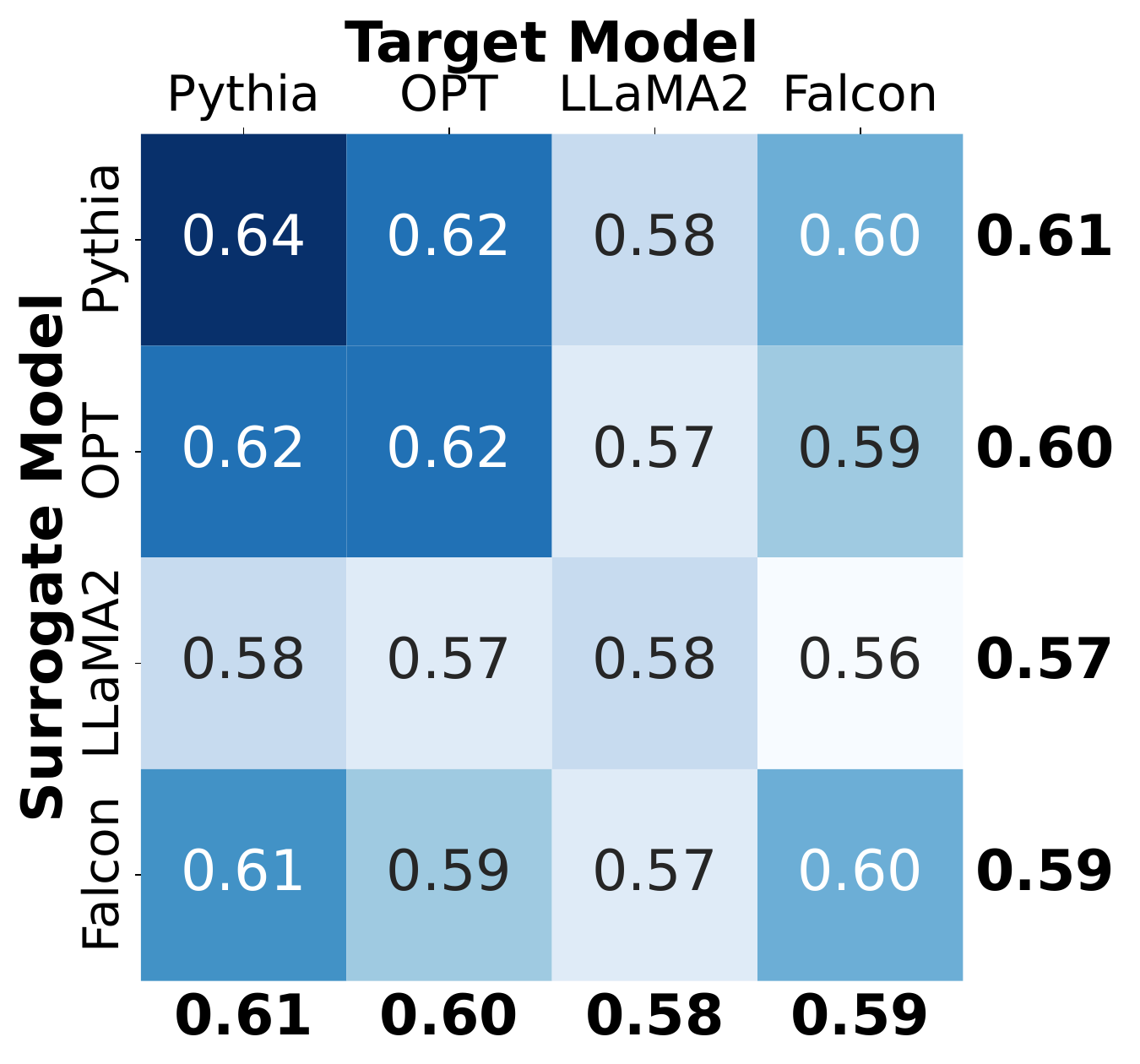}
    \caption{Results of PETAL with various surrogate models.}
    \vspace{-0.5em}
    \label{Figure: ablation on surrogate model}
\end{figure}

\partitle{Different Surrogate Model}
Because of the differences in properties between the surrogate and target models, there may be errors between the approximated output probabilities and the actual probabilities. The choice of the surrogate model could thus affect the attack performance. We explore this influence by evaluating attacks under various target model-surrogate model combinations, and the results are presented in Figure \ref{Figure: ablation on surrogate model}. It demonstrates the attack performance remains relatively stable despite variations in combinations (particularly, AUC changes within 0.06), proving the robustness of PETAL. These results align with our detailed analysis in Section \ref{sec: 4.3}, further confirming that PETAL's effectiveness stems from the use of token-level semantic similarity itself. Nevertheless, since there are subtle performance differences, future work could investigate how to utilize the surrogate model to enhance label-only MIAs.    

\begin{figure}[t]
    \centering
    \includegraphics[width=0.44\textwidth]{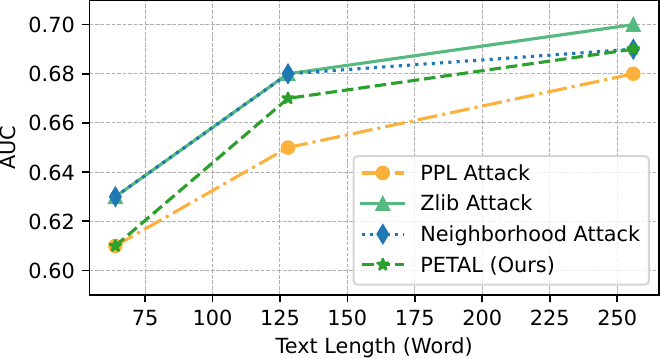}
    \caption{AUC values of various attacks against Pythia-6.9B. The text length is set at 64, 128, and 256 words.}
    \label{Figure: as_length}
\vspace{-1.5em}
\end{figure}

\partitle{Different Text Length} 
We evaluate the performance of various attacks under different text length settings (64, 128, 256). The results in Figure \ref{Figure: as_length} indicate a general improvement in the performance of all attacks as the text length increases. That can be attributed to longer texts containing more memorized prefix-token pairs, making members easier to identify. Therefore controlling the maximum length of user prompts could likely mitigate the privacy leakage caused by MIAs. On the other hand, PETAL exhibits comparable performance to other logits-based attacks across different text lengths, demonstrating its robustness again.

\begin{table}[!t]
\newcommand{\tabincell}[2]{\begin{tabular}{@{}#1@{}}#2\end{tabular}}
\centering
\setlength{\tabcolsep}{2.5pt}
\caption{The proportion of parameter sets that result in an improvement in AUC. The dataset is WikiMIA \cite{shi2023detecting}.}
\vspace{-1mm}
\scalebox{1.0}
{
\begin{tabular}{c|ccccc}
\toprule
AUC boost&Pythia &OPT &LLaMA2 &Falcon &\textbf{ALL}\\
\midrule
$\geq0.01$& 5.3\%& 3.7\%& 7.7\%& 4.2\% &\textbf{0.0\%}\\
$\geq0.02$& 1.6\%& 0.4\%& 1.9\%& 0.4\% &\textbf{0.0\%}\\
$\geq0.03$& 0.3\%& 0.0\%& 0.3\%& 0.0\% &\textbf{0.0\%}\\
$\geq0.04$& 0.0\%& 0.0\%& 0.0\%& 0.0\% &\textbf{0.0\%}\\
\bottomrule
\end{tabular}
}
\vspace{-0.5em}
\label{table: ablation on weighting}
\end{table}

\begin{table}[!t]
\footnotesize
\newcommand{\tabincell}[2]{\begin{tabular}{@{}#1@{}}#2\end{tabular}}
\centering
\setlength{\tabcolsep}{2.5pt}
\caption{Performance comparison of PETAL under different sentence transformer settings on WikiMIA \cite{shi2023detecting}.}
\vspace{-1mm}
\scalebox{1.0}
{
\begin{tabular}{c|ccccc}
\toprule
Sentence transformers&TPR@1\%FPR$\uparrow$ &AUC$\uparrow$ &Balanced Acc$\uparrow$\\
\midrule
all-MiniLM-L6-v2    & 4.9\%& 0.64& 0.62\\
bge-large-en-v1.5   & 3.9\%& 0.64& 0.62\\
UAE-Large-V1        & 4.1\%& 0.65& 0.62\\
mxbai-embed-large-v1& 5.4\%& 0.64& 0.61\\
\bottomrule
\end{tabular}
}
\vspace{-0.5em}
\label{table: ablation on sentence transformers}
\end{table}

\partitle{Different Decoding Strategy}
As we focus on the label-only settings, decoding strategies might influence attack performance. To explore whether PETAL is sensitive to different decoding strategies, we evaluate it under three representative strategies including greedy search, nucleus sampling \cite{holtzman2019curious}, and contrastive search \cite{su2022a}. Specifically, greedy search selects the token with the highest probability as its next token. Nucleus sampling chooses from the smallest possible set of tokens whose cumulative probability exceeds a pre-designed threshold $p$. The probability mass is then redistributed among this set of tokens. Contrastive search jointly considers the probability predicted by the language model and the similarity with respect to the previous context (to prevent model degradation). For nucleus sampling, we set $p=0.95$, and for contrastive search, we set $k=4$ (controls the size of the candidate set) and $\alpha=0.6$ (regulates the importance of model confidence and degradation penalty). Table \ref{table: as_decoding} demonstrates that PETAL is stable and not sensitive to decoding strategies. The adversary can thus conduct an effective attack against LLMs without knowledge of decoding strategies. On the other hand, since PETAL is built upon the principle that the probability assigned to the generated token is high, the AUC results exhibit a marginal increase when the sampling strategy is set to greedy search (where the probability assigned to the generated token is highest).

\partitle{Different Weighting Strategy}
Since the relationship between probability and semantic similarity may vary from token to token depending on their position in the sentence, employing more sophisticated weighting strategies could potentially bring improvements. To investigate this, we randomly sample 100,000 sets of parameters from the Dirichlet distribution (this ensures the parameters sum to 1) and calculate the corresponding AUC boost compared to unweighted average-based perplexity. Table \ref{table: ablation on weighting} shows that no parameter set outperforms unweighted perplexity across all 4 models, although slight improvements are observed in about 5\% of cases for individual models. We note that it is challenging for attackers to identify parameter sets specific to target LLMs in real-world scenarios, due to the lack of ground truth. We will further explore its designs in our future work.

\partitle{Different Sentence Transformers}
We hereby explore whether our PETAL is still effective under different sentence transformers. Specifically, we use multiple sentence transformers released between 2021 and 2024, including all-MiniLM-L6-v2 \cite{lewis2020bart}, bge-large-en-v1.5 \cite{bge_embedding}, UAE-Large-V1 \cite{li2023angle}, and mxbai-embed-large-v1 \cite{emb2024mxbai}, for discussion. The results in Table \ref{table: ablation on sentence transformers} show that PETAL is robust to various sentence transformers, despite their different characteristics arising from variations in architecture and training data.

\begin{table}[!t]
\footnotesize
\newcommand{\tabincell}[2]{\begin{tabular}{@{}#1@{}}#2\end{tabular}}
\centering
\setlength{\tabcolsep}{2.5pt}
\caption{Attack results on LLaMA-Doctor and OPT-History show that PETAL remains effective on fine-tuned LLMs.}
\vspace{-1mm}
\scalebox{0.96}
{
\begin{tabular}{c|cccccc}
\toprule
Metrics & \multicolumn{2}{c}{TPR@1\%FPR$\uparrow$}& \multicolumn{2}{c}{AUC$\uparrow$}& \multicolumn{2}{c}{Balanced Acc$\uparrow$}\\
\cmidrule(l{5pt}r{5pt}){2-3}\cmidrule(l{5pt}r{5pt}){4-5}\cmidrule(l{5pt}r{5pt}){6-7}
Models & LLaMA-3.2& OPT& LLaMA-3.2& OPT& LLaMA-3.2& OPT\\
\midrule
PPL attack& 19.2\%& 7.8\%& 0.86& 0.75& 0.78& 0.70\\
WS attack& 13.0\% & 1.3\%& 0.65& 0.50& 0.64& 0.51\\
RS attack& 3.5\%  & 1.8\%& 0.54& 0.50& 0.52& 0.51\\
BT attack& 2.7\%  & 1.3\%& 0.59& 0.49& 0.58& 0.51\\
PETAL (Ours)& \textbf{20.7\%}& \textbf{8.9\%}& \textbf{0.87}& \textbf{0.74}& \textbf{0.79}& \textbf{0.68}\\
\bottomrule
\end{tabular}
}
\vspace{-1em}
\label{table: fine-tuning results on chatdoctor}
\end{table}

\begin{table*}[!t]
\centering
\setlength{\tabcolsep}{4.0pt}
\caption{AUC and TPR at 1\% FPR results of various attacks on WikiMIA when the input text is paraphrased. Values in parentheses indicate the difference in attack performance between paraphrased and non-paraphrased text.}
\scalebox{0.83}
{
\begin{tabular}{l|cccccccc}
\toprule
Metrics & \multicolumn{4}{c}{AUC $\uparrow$}&  \multicolumn{4}{c}{TPR@1\%FPR  $\uparrow$}\\
\cmidrule(l{5pt}r{5pt}){2-5}\cmidrule(l{5pt}r{5pt}){6-9}
Models & Pythia-6.9B& OPT-6.7B& Falcon-7B& LLaMA2-13B& Pythia-6.9B& OPT-6.7B& Falcon-7B& LLaMA2-13B\\
\midrule
\textbf{\textit{Logits-based Attacks}}&&&&&&&&\\
PPL attack&0.64 (+0.00)&0.60 (-0.00)&0.55 (+0.00)&0.59 (+0.00)&6.2\% (+0.0\%)&2.6\% (-0.8\%)&1.3\% (+0.0\%)&2.1\% (+1.0\%)\\
reference attack&0.51 (+0.01)&0.44 (-0.00)&0.40 (-0.00)&0.44 (+0.01)&1.8\% (-0.3\%)&0.3\% (+0.0\%)&0.0\% (+0.0\%)&0.3\% (+0.3\%)\\
zlib attack&0.64 (-0.00)&0.61 (-0.00)&0.56 (+0.00)&0.60 (+0.00)&5.2\% (+0.3\%)&2.3\% (-1.0\%)&1.6\% (-0.8\%)&2.8\% (-0.5\%)\\
neighborhood attack&0.66 (-0.00)&0.63 (-0.01)&0.56 (-0.01)&0.61 (-0.01)&0.8\% (+0.0\%)&1.8\% (+1.0\%)&0.3\% (-0.3\%)&0.8\% (-1.8\%)\\
MIN-K\% PROB&0.65 (-0.01)&0.60 (-0.01)&0.53 (-0.00)&0.56 (-0.01)&6.7\% (-2.1\%)&2.6\% (-1.8\%)&1.6\% (+0.8\%)&3.4\% (-0.8\%)\\
\textbf{\textit{Label-only Attacks}}&&&&&&&&\\
PETAL (Ours)&0.64 (+0.00)&0.62 (-0.00)&0.58 (-0.00)&0.60 (-0.00)&2.1\% (-2.8\%)&1.8\% (-1.3\%)&1.6\% (+0.3\%)&1.8\% (-0.3\%)\\
\bottomrule
\end{tabular}
}

\label{table:as_paraphrasing}
\end{table*}

\begin{table*}[!t]
\centering
\setlength{\tabcolsep}{4.0pt}
\caption{AUC results of various attacks on two subsets of MIMIR when the target model is pre-trained on deduplicated training data. Values in parentheses indicate the difference in attack performance between the deduped and non-deduped target model.}
\scalebox{0.88}
{
\begin{tabular}{l|cccccccc}
\toprule
Subsets & \multicolumn{4}{c}{GitHub}&  \multicolumn{4}{c}{HackerNews}\\
\cmidrule(l{5pt}r{5pt}){2-5}\cmidrule(l{5pt}r{5pt}){6-9}
Models & 160M& 1.4B& 2.8B& 6.9B& 160M& 1.4B& 2.8B& 6.9B\\
\midrule
\textbf{\textit{Logits-based Attacks}}&&&&&&&&\\
PPL attack&0.84 (-0.00)&0.86 (-0.00)&0.87 (-0.00)&0.87 (-0.01)&0.59 (-0.00)&0.59 (-0.00)&0.60 (-0.00)&0.59 (-0.01)\\
reference attack&0.64 (-0.00)&0.62 (-0.03)&0.64 (-0.01)&0.61 (-0.03)&0.53 (+0.00)&0.52 (-0.00)&0.53 (+0.00)&0.52 (-0.00)\\
zlib attack&0.83 (-0.00)&0.85 (-0.00)&0.86 (-0.00)&0.86 (-0.01)&0.59 (+0.00)&0.59 (-0.00)&0.59 (-0.00)&0.59 (-0.00)\\
neighborhood attack&0.83 (+0.03)&0.83 (+0.00)&0.82 (-0.00)&0.82 (-0.02)&0.53 (-0.03)&0.54 (+0.00)&0.54 (-0.00)&0.54 (-0.01)\\
MIN-K\% PROB&0.80 (-0.03)&0.85 (-0.01)&0.86 (-0.00)&0.87 (-0.01)&0.56 (-0.01)&0.56 (-0.01)&0.58 (-0.00)&0.56 (-0.02)\\
\textbf{\textit{Label-only Attacks}}&&&&&&&&\\
PETAL (Ours)&0.83 (-0.00)&0.85 (-0.00)&0.86 (-0.00)&0.86 (-0.01)&0.59 (-0.00)&0.58 (-0.01)&0.58 (-0.00)&0.59 (+0.01)\\
\bottomrule
\end{tabular}
}
\vspace{-0.5em}
\label{table:as_deduplicaion}
\end{table*}

\subsection{Results on Fine-tuned LLMs}
Although we focus on the pre-training phase because it is more prevalent and difficult, it is also important to explore whether our PETAL is also effective in identifying members from fine-tuning datasets or requires additional adjustments.

\partitle{Attack Setup}
To answer these questions, we evaluate PETAL on LLaMA-Doctor\footnote{\url{https://huggingface.co/prithivMLmods/Llama-Doctor-3.2-3B-Instruct}} and OPT-History\footnote{\url{https://huggingface.co/ambrosfitz/opt-history-v2}}, which are fine-tuned versions of LLaMA-3.2-3B-Instruct and OPT-350M on Chatdoctor\footnote{\url{https://huggingface.co/datasets/avaliev/chat_doctor}} and History\footnote{\url{https://huggingface.co/datasets/ambrosfitz/just_history}} datasets, respectively. Note that the two models we attack are fine-tuned by developers on Hugging Face, not by us, to align with real-world scenarios.

\partitle{Attack Results}
The results in Table \ref{table: fine-tuning results on chatdoctor} show that: 1) MIAs pose a more severe threat on the fine-tuning phase compared to the pre-training phase (e.g., the AUCs can even reach 0.87 on LLaMA-Doctor); 2) The performance of PETAL remains on par with logits-based attacks and significantly outperforms other label-only attacks. This is not surprising\textemdash the main difference between the fine-tuning and pre-training phases lies in the number of times the model exposes the training samples. The generalization gap between members and non-members still exists, and potentially becomes larger.

\vspace{-0.5em}
\section{PETAL against Defenses} 

While defense against MIAs is not the primary focus of this work, we explore several well-established defense strategies for simple models and LLMs, including deduplication, MemGuard \cite{jia2019memguard}, text paraphrasing, and differential privacy \cite{dwork2006calibrating}. We discuss their effectiveness in mitigating PETAL in detail.

\partitle{Text Paraphrasing}
Text paraphrasing is a data augmentation technique in NLP. Since \textit{paraphrased members} are semantically highly similar to the original ones and can also reveal their existence, the current community generally agrees that MIAs against LLMs should infer \textit{paraphrased members} as members \cite{shi2023detecting}. To investigate its defensive effect, we use the paraphrased WikiMIA benchmark provided by Zhang et al. \cite{zhang2024min} to evaluate the performance of various attacks on four LLMs. The results in Table \ref{table:as_paraphrasing} indicate that even when the input text is paraphrased, PETAL's performance shows little to no decline, especially in terms of AUC. This is likely because PETAL utilizes semantic information contained in tokens, which is hardly altered by text paraphrasing. Additional results of balanced accuracy can be found in Appendix \ref{sec: balanced acc on wiki-para}.

\partitle{Training Data Deduplication}
Studies have shown that data deduplication could mitigate the privacy risks introduced by MIAs in pre-trained LLMs \cite{lee2022deduplicating,kandpal2022deduplicating}. This approach is more efficient than training-based methods (e.g., differential privacy) and is likely a practical solution. To validate the effect of data deduplication on our PETAL, we use the Pythia-dedup family \cite{biderman2023pythia} as the target model, which has the same parameter counts as the Pythia family but is trained on the deduplicated Pile. The results in Table \ref{table:as_deduplicaion} indicate that data deduplication only slightly reduces MIA performance (to an almost negligible extent). Our findings suggest that coarse-grained deduplication is insufficient to effectively mitigate LLM memorization, which aligns with prior studies \cite{nasr2023scalable}. However, finer-grained deduplication might remove useful knowledge \cite{ishihara2023training}, and deduplication itself requires retraining the model, which is often unaffordable for typical companies. We believe that more effective countermeasures are needed to address the risks posed by MIAs. Additional results are in Appendix \ref{sec: additional results on granularity}.

\partitle{MemGuard}
MemGuard \cite{jia2019memguard} is a confidence score masking-based defense that protects member privacy by hiding the true probabilities output by the target models. Specifically, it adds a carefully crafted noise to probability vectors assigned by the target model. The carefully crafted noise could turn the probability vectors into an adversarial example, which will mislead the adversary to make a random guess. Although MemGuard has shown effectiveness against logits-based attacks \cite{choquette2021label}, it is ineffective against our PETAL. This is because PETAL does not use the model's actual output probabilities but instead approximates probabilities using semantic similarity. Specifically, as MemGuard guarantees that the crafted noise does not change the predicted label for any query (to bound utility loss), PETAL can still compute accurate semantic similarity and approximate the actual probability vectors accordingly.

\partitle{Differential Privacy}
Differential privacy \cite{dwork2006calibrating} is a widely used defense mechanism against MIAs. It imposes theoretical bounds on the success rate of MIAs by directly restricting the ability to distinguish between two neighboring datasets (i.e., differing only in the inclusion or exclusion of a particular sample). While differential privacy is the gold standard for defending against MIAs, applying classic algorithms like DPSGD \cite{abadi2016deep} to the pre-training phase of LLMs is challenging---increased computation and memory usage for re-pre-training an LLM (e.g., 7B-level) with differential privacy is prohibitive for researchers. To the best of our knowledge, existing work \cite{heexploring} focuses on how to privately fine-tune an LLM, and currently there are no publicly available instances of LLMs that are pre-trained with differential privacy. We hope to evaluate PETAL on differentially private LLMs in the future if possible.

\vspace{-0.5em}
\section{Case Studies on Closed-source LLMs}
We observe that previous research on membership inference attacks has not explored closed-source large language models. This may be due to: 1) many LLMs APIs not releasing full logits or limiting outputs to the top-$k$ logits for each query, and 2) model owners' reluctance to disclose sources and techniques of pre-training data collection to protect their intellectual property. To address this gap, we conduct two case studies on Gemini-1.5-Flash \cite{team2023gemini} and GPT-3.5-Turbo-Instruct \cite{achiam2023gpt}.

\partitle{Attack Setup}
To maintain a competitive edge, model owners are increasingly disclosing less information about the sources of their pre-training data in technical reports or blogs. For instance, Anthropic describes Claude's pre-training data merely as ``a proprietary mix of publicly available information on the Internet'' \cite{anthropic2024claude}. To evaluate the effectiveness of PETAL on commercial closed-source LLMs, we use Gemini-1.5-Flash and GPT-3.5-Turbo-Instruct as target models, with the updated version of WikiMIA~\cite{fu2024mia} as the target dataset. It uses March 2024 as the cutoff date for distinguishing members from non-members. Note that our setup may not be perfect due to the lack of detailed knowledge about the training data for these two LLMs. However, it is somewhat reasonable and generally sufficient for validating the effectiveness of PETAL on real-world LLMs (see more details in Appendix \ref{sec: Case_study_setup}).

\partitle{Attack Results} 
Table \ref{table: case studies on commercial LLMs} demonstrates that PETAL remains effective to advanced commercial LLMs, which for the first time reveals the privacy threats posed by MIAs in real-world scenarios\textemdash without any access to model parameters or output probability distributions, PETAL achieves an attack AUC of 0.67 on GPT-3.5-Turbo-Instruct. Another interesting finding is that Gemini-1.5 appears to have weaker memorization compared to GPT-3.5-Turbo-Instruct. One possible reason is that the additional safety alignment the former underwent mitigates the model's leakage of training data \cite{nasr2023scalable}.

\begin{table}[!t]
\footnotesize
\newcommand{\tabincell}[2]{\begin{tabular}{@{}#1@{}}#2\end{tabular}}
\centering
\setlength{\tabcolsep}{2.5pt}
\caption{Attack results on Gemini-1.5 and GPT-3.5-Turbo-Instruct. The dataset is an updated version of WikiMIA \cite{fu2024mia}.}
\scalebox{1.0}
{
\begin{tabular}{c|cccccc}
\toprule
Metrics & \multicolumn{2}{c}{TPR@1\%FPR$\uparrow$}& \multicolumn{2}{c}{AUC$\uparrow$}& \multicolumn{2}{c}{Balanced Acc$\uparrow$}\\
\cmidrule(l{5pt}r{5pt}){2-3}\cmidrule(l{5pt}r{5pt}){4-5}\cmidrule(l{5pt}r{5pt}){6-7}
Models & GPT-3.5& Gemini& GPT-3.5& Gemini& GPT-3.5& Gemini\\
\midrule
PETAL (Ours)& 4.2\%& 2.5\%& 0.67& 0.61& 0.64& 0.58\\
\bottomrule
\end{tabular}
}
\vspace{-1em}
\label{table: case studies on commercial LLMs}
\end{table}

\vspace{-0.5em}
\section{Discussion}

\partitle{Efficiency and Practicality}
We acknowledge that as a label-only MIA, PETAL also encounters traditional efficiency limitations. However, we argue that from the perspective of query cost, PETAL is still efficient. Specifically, recall that implementing PETAL on a $n$-token long sample costs $n(n-1)/2$ input tokens and $n-1$ output tokens, which add up to approximately $(n-1)(n+2)/2$ input tokens. Based on our analysis in Section \ref{sec: 5.3}, this budget can be further reduced to $(n-1)(n+2)/8$ input tokens. Taking GPT-3.5-Turbo-Instruct as an example, attacking a 128-token long sample costs only \$0.003. Besides, the performance of PETAL is on par with that of logits-based attacks without any access to output probabilities. In summary, PETAL is practical due to its acceptable attack cost and reduced demands on attacker capabilities.

\partitle{Theoretical Underpinnings}
Distributed representations, as the foundational concept underlying modern LLMs naturally lead to the following phenomenon: the presence of one sentence in the training corpus increases both the probability of that sentence and its similar ``neighbors'' in semantic space \cite{bengio2000neural}. Consequently, well-trained LLMs tend to assign similar output probabilities to semantically related tokens. Besides, since the model has assigned a high probability to the generated token (otherwise it would not have been sampled), higher token-level semantic similarity naturally implies a higher output probability (\ie, stronger MIA signals). We will explore the theory behind PETAL's success in future work.

\partitle{Limitations}
We admit that our work still has some potential limitations. Firstly, our method requires the use of a surrogate model. Although there are many easily accessible open-source LLMs that could serve as the surrogate model, deploying an additional LLM demands more GPU resources. We will explore how to extend our method in the `surrogate-free' cases in our future works. Secondly, similar to existing work~\cite{ko2023practical,shi2023detecting,zhang2024min}, this paper mainly focuses on the privacy leakage risk posed by MIAs. Other possible downstream applications such as auditing machine unlearning~\cite{bourtoule2021machine} and detecting copyright infringement~\cite{shao2025explanation} are left for our future work. Despite these limitations, we believe our study provides significant insights into label-only MIAs against large language models. 

\vspace{-0.5em}
\section{Conclusion}
In this paper, we investigate whether LLMs are vulnerable to MIAs in the label-only setting. We focus on the pre-training phase as it is more prevalent and presents greater technical challenges. We first summarize representative MIAs designed for LMs and reveal that existing label-only attacks are ineffective to pre-trained LLMs. To fill this gap, we propose PETAL: a label-only membership inference attack based on per-token semantic similarity. Specifically, PETAL leverages token-level semantic similarity to approximate output probabilities and subsequently calculate the perplexity. Experimental results across various metrics demonstrate the effectiveness of our method. Our research uncovers that pre-trained LLMs suffer from equivalent privacy risk even in the most stringent label-only settings. We hope that our analysis can motivate the community to develop more effective defenses.

\section*{Acknowledgments}
This research is supported in part by the National Key Research and Development Program of China under Grant 2021YFB3100300, the National Natural Science Foundation of China under Grants (62441238, 62072395, and U20A20178), the Zhejiang Provincial Natural Science Foundation of China under Grant LD24F020014, and the NTU-NAP startup grant. This work was mostly done when Yiming Li was a research professor at the State Key Laboratory of Blockchain and Data Security, Zhejiang University. He is currently at Nanyang Technological University.

\section*{Ethics Considerations}

As is the case with any examination of the security vulnerabilities present in LLMs, it is imperative to emphasize that we do not endorse the practical application of such attacks. Our aim in presenting this study is to caution the LLM community, highlighting the inherent risks of MIAs even within the most stringent attack settings. We underscore the necessity of deepening our understanding of LLM memorization mechanisms and the development of robust defense strategies, particularly during the pre-training phase. However, due to potential privacy risks associated with our paper, we have shared our findings with the authors of all the open-source models we discuss in this paper (e.g., Pythia \cite{biderman2023pythia}, OPT \cite{zhang2022opt}, LLaMA2 \cite{touvron2023llamab} and Falcon \cite{almazrouei2023falcon}). We have also disclosed the vulnerability of the two commercial LLMs studied in this paper to OpenAI and Google respectively, providing them with a 60-day period to address the issue before publishing this paper. Despite our efforts, considering that the memorization of training data by LLMs may be an inherent property and that the issue of MIAs is difficult to fully resolve, we strongly encourage developers to avoid including privacy-sensitive data in pre-training datasets without implementing robust defensive measures.

In regard to the data utilized, they are all publicly available and have been extensively used in prior research, so there are no privacy concerns. We use data from WikiMIA and MIMIR for only research proposes and this wouldn't violate the license in most cases.

\section*{Open Science}

Our research team is dedicated to upholding open science principles by making our findings freely accessible. This commitment extends to sharing all research-related materials, including datasets, scripts, and source code, to foster a wider adoption of open science practices.

\partitle{Open sharing of code and other resources}
To facilitate academic collaboration and technological progress, we intend to publish all research artifacts on GitHub, making them publicly available. This includes datasets, scripts, and source code used in our study. It's worth noting that the models employed in our main experiments (e.g., Pythia, LLaMA, OPT, etc.) are open-source and can be freely accessed and downloaded online (e.g., on Hugging Face\footnote{\url{https://huggingface.co/models}}). Gemini-1.5-Flash can be accessed using the official API and technical documentation provided by the Google AI team\footnote{\url{https://ai.google.dev/gemini-api/docs}}. GPT-3.5-Turbo-Instruct can also be accessed using the official API and technical documentation provided by OpenAI\footnote{\url{https://platform.openai.com/docs/guides/text-generation}}. Regarding datasets, this paper mainly uses WikiMIA benchmark \cite{shi2023detecting} and MIMIR benchmark \cite{duan2024membership}, both of which are openly accessible on the Hugging Face platform, ensuring transparency. Both the WikiMIA and MIMIR benchmarks are released under the MIT license.

\partitle{Reproducibility and Replicability}
To ensure the reproducibility of our work, all artifacts necessary for replicating the results presented in our paper will be meticulously documented and made publicly available in our GitHub repository. This comprehensive documentation includes but is not limited to, detailed environment configurations, source code, hyperparameter settings, and other pertinent experimental details. By providing this level of transparency, we aim to establish a verifiable foundation for our research, facilitating further scientific exploration and validation within the academic and research communities.

{\footnotesize \bibliographystyle{style/acm}
\bibliography{ref}}

\begin{thebibliography}{10}

\bibitem{abadi2016deep}
{\sc Abadi, M., Chu, A., Goodfellow, I., McMahan, H.~B., Mironov, I., Talwar, K., and Zhang, L.}
\newblock Deep learning with differential privacy.
\newblock In {\em CCS\/} (2016), pp.~308--318.

\bibitem{achiam2023gpt}
{\sc Achiam, J., Adler, S., Agarwal, S., Ahmad, L., Akkaya, I., Aleman, F.~L., Almeida, D., Altenschmidt, J., Altman, S., Anadkat, S., et~al.}
\newblock Gpt-4 technical report.
\newblock {\em arXiv preprint arXiv:2303.08774\/} (2023).

\bibitem{almazrouei2023falcon}
{\sc Almazrouei, E., Alobeidli, H., Alshamsi, A., Cappelli, A., Cojocaru, R., Debbah, M., Goffinet, E., Heslow, D., Launay, J., Malartic, Q., et~al.}
\newblock Falcon-40b: an open large language model with state-of-the-art performance.
\newblock {\em Findings of the Association for Computational Linguistics: ACL 2023\/} (2023), 10755--10773.

\bibitem{anthropic2024claude}
{\sc Anthropic, A.}
\newblock The claude 3 model family: Opus, sonnet, haiku.
\newblock {\em Claude-3 Model Card 1\/} (2024).

\bibitem{bai2023qwen}
{\sc Bai, J., Bai, S., Chu, Y., Cui, Z., Dang, K., Deng, X., Fan, Y., Ge, W., Han, Y., Huang, F., et~al.}
\newblock Qwen technical report.
\newblock {\em arXiv preprint arXiv:2309.16609\/} (2023).

\bibitem{bengio2000neural}
{\sc Bengio, Y., Ducharme, R., and Vincent, P.}
\newblock A neural probabilistic language model.
\newblock {\em NeurIPS 13\/} (2000).

\bibitem{biderman2023pythia}
{\sc Biderman, S., Schoelkopf, H., Anthony, Q.~G., Bradley, H., O’Brien, K., Hallahan, E., Khan, M.~A., Purohit, S., Prashanth, U.~S., Raff, E., et~al.}
\newblock Pythia: A suite for analyzing large language models across training and scaling.
\newblock In {\em ICML\/} (2023), PMLR, pp.~2397--2430.

\bibitem{bourtoule2021machine}
{\sc Bourtoule, L., Chandrasekaran, V., Choquette-Choo, C.~A., Jia, H., Travers, A., Zhang, B., Lie, D., and Papernot, N.}
\newblock Machine unlearning.
\newblock In {\em 2021 IEEE Symposium on Security and Privacy (SP)\/} (2021), IEEE, pp.~141--159.

\bibitem{brown2020language}
{\sc Brown, T., Mann, B., Ryder, N., Subbiah, M., Kaplan, J.~D., Dhariwal, P., Neelakantan, A., Shyam, P., Sastry, G., Askell, A., et~al.}
\newblock Language models are few-shot learners.
\newblock {\em NeurIPS 33\/} (2020), 1877--1901.

\bibitem{bucknall2023structured}
{\sc Bucknall, B.~S., and Trager, R.~F.}
\newblock Structured access for third-party research on frontier ai models: Investigating researchers’model access requirements, 2023.

\bibitem{carlini2022membership}
{\sc Carlini, N., Chien, S., Nasr, M., Song, S., Terzis, A., and Tramer, F.}
\newblock Membership inference attacks from first principles.
\newblock In {\em 2022 IEEE Symposium on Security and Privacy (SP)\/} (2022), IEEE, pp.~1897--1914.

\bibitem{carlini2021extracting}
{\sc Carlini, N., Tramer, F., Wallace, E., Jagielski, M., Herbert-Voss, A., Lee, K., Roberts, A., Brown, T., Song, D., Erlingsson, U., et~al.}
\newblock Extracting training data from large language models.
\newblock In {\em 30th USENIX Security Symposium (USENIX Security 21)\/} (2021), pp.~2633--2650.

\bibitem{choquette2021label}
{\sc Choquette-Choo, C.~A., Tramer, F., Carlini, N., and Papernot, N.}
\newblock Label-only membership inference attacks.
\newblock In {\em ICML\/} (2021), PMLR, pp.~1964--1974.

\bibitem{chowdhery2023palm}
{\sc Chowdhery, A., Narang, S., Devlin, J., Bosma, M., Mishra, G., Roberts, A., Barham, P., Chung, H.~W., Sutton, C., Gehrmann, S., et~al.}
\newblock Palm: Scaling language modeling with pathways.
\newblock {\em Journal of Machine Learning Research 24}, 240 (2023), 1--113.

\bibitem{duan2024membership}
{\sc Duan, M., Suri, A., Mireshghallah, N., Min, S., Shi, W., Zettlemoyer, L., Tsvetkov, Y., Choi, Y., Evans, D., and Hajishirzi, H.}
\newblock Do membership inference attacks work on large language models?
\newblock {\em arXiv preprint arXiv:2402.07841\/} (2024).

\bibitem{dwork2006calibrating}
{\sc Dwork, C., McSherry, F., Nissim, K., and Smith, A.}
\newblock Calibrating noise to sensitivity in private data analysis.
\newblock In {\em Theory of Cryptography: Third Theory of Cryptography Conference, TCC 2006, New York, NY, USA, March 4-7, 2006. Proceedings 3\/} (2006), Springer, pp.~265--284.

\bibitem{fu2023practical}
{\sc Fu, W., Wang, H., Gao, C., Liu, G., Li, Y., and Jiang, T.}
\newblock Practical membership inference attacks against fine-tuned large language models via self-prompt calibration.
\newblock {\em arXiv preprint arXiv:2311.06062\/} (2023).

\bibitem{fu2024mia}
{\sc Fu, W., Wang, H., Gao, C., Liu, G., Li, Y., and Jiang, T.}
\newblock Mia-tuner: Adapting large language models as pre-training text detector.
\newblock {\em arXiv preprint arXiv:2408.08661\/} (2024).

\bibitem{gao2020pile}
{\sc Gao, L., Biderman, S., Black, S., Golding, L., Hoppe, T., Foster, C., Phang, J., He, H., Thite, A., Nabeshima, N., et~al.}
\newblock The pile: An 800gb dataset of diverse text for language modeling.
\newblock {\em arXiv preprint arXiv:2101.00027\/} (2020).

\bibitem{gilbert2023large}
{\sc Gilbert, S., Harvey, H., Melvin, T., Vollebregt, E., and Wicks, P.}
\newblock Large language model ai chatbots require approval as medical devices.
\newblock {\em Nature Medicine 29}, 10 (2023), 2396--2398.

\bibitem{ginart2019making}
{\sc Ginart, A., Guan, M., Valiant, G., and Zou, J.~Y.}
\newblock Making ai forget you: Data deletion in machine learning.
\newblock {\em NeurIPS 32\/} (2019).

\bibitem{heexploring}
{\sc He, J., Li, X., Yu, D., Zhang, H., Kulkarni, J., Lee, Y.~T., Backurs, A., Yu, N., and Bian, J.}
\newblock Exploring the limits of differentially private deep learning with group-wise clipping.
\newblock In {\em ICLR\/} (2023).

\bibitem{he2024difficulty}
{\sc He, Y., Li, B., Wang, Y., Yang, M., Wang, J., Hu, H., and Zhao, X.}
\newblock Is difficulty calibration all we need? towards more practical membership inference attacks.
\newblock In {\em CCS\/} (2024), pp.~1226--1240.

\bibitem{hoang2019efficient}
{\sc Hoang, A., Bosselut, A., Celikyilmaz, A., and Choi, Y.}
\newblock Efficient adaptation of pretrained transformers for abstractive summarization.
\newblock {\em arXiv preprint arXiv:1906.00138\/} (2019).

\bibitem{holtzman2019curious}
{\sc Holtzman, A., Buys, J., Du, L., Forbes, M., and Choi, Y.}
\newblock The curious case of neural text degeneration.
\newblock In {\em ICLR\/} (2020).

\bibitem{hu2019new}
{\sc Hu, S., Yu, T., Guo, C., Chao, W.-L., and Weinberger, K.~Q.}
\newblock A new defense against adversarial images: Turning a weakness into a strength.
\newblock {\em NeurIPS 32\/} (2019).

\bibitem{ishihara2023training}
{\sc Ishihara, S.}
\newblock Training data extraction from pre-trained language models: A survey.
\newblock In {\em The Third Workshop on Trustworthy Natural Language Processing\/} (2023), p.~260.

\bibitem{jia2019memguard}
{\sc Jia, J., Salem, A., Backes, M., Zhang, Y., and Gong, N.~Z.}
\newblock Memguard: Defending against black-box membership inference attacks via adversarial examples.
\newblock In {\em CCS\/} (2019), pp.~259--274.

\bibitem{kandpal2022deduplicating}
{\sc Kandpal, N., Wallace, E., and Raffel, C.}
\newblock Deduplicating training data mitigates privacy risks in language models.
\newblock In {\em ICML\/} (2022), PMLR, pp.~10697--10707.

\bibitem{kenton2019bert}
{\sc Kenton, J. D. M.-W.~C., and Toutanova, L.~K.}
\newblock Bert: Pre-training of deep bidirectional transformers for language understanding.
\newblock In {\em Proceedings of naacL-HLT\/} (2019), vol.~1, p.~2.

\bibitem{ko2023practical}
{\sc Ko, M., Jin, M., Wang, C., and Jia, R.}
\newblock Practical membership inference attacks against large-scale multi-modal models: A pilot study.
\newblock In {\em Proceedings of the IEEE/CVF International Conference on Computer Vision\/} (2023), pp.~4871--4881.

\bibitem{lee2023language}
{\sc Lee, J., Le, T., Chen, J., and Lee, D.}
\newblock Do language models plagiarize?
\newblock In {\em Proceedings of the ACM Web Conference 2023\/} (2023), pp.~3637--3647.

\bibitem{lee2022deduplicating}
{\sc Lee, K., Ippolito, D., Nystrom, A., Zhang, C., Eck, D., Callison-Burch, C., and Carlini, N.}
\newblock Deduplicating training data makes language models better.
\newblock In {\em ACL\/} (2022).

\bibitem{emb2024mxbai}
{\sc Lee, S., Shakir, A., Koenig, D., and Lipp, J.}
\newblock Open source strikes bread - new fluffy embeddings model, 2024.

\bibitem{leino2020stolen}
{\sc Leino, K., and Fredrikson, M.}
\newblock Stolen memories: Leveraging model memorization for calibrated $\{$White-Box$\}$ membership inference.
\newblock In {\em 29th USENIX security symposium (USENIX Security 20)\/} (2020), pp.~1605--1622.

\bibitem{lewis2020bart}
{\sc Lewis, M., Liu, Y., Goyal, N., Ghazvininejad, M., Mohamed, A., Levy, O., Stoyanov, V., and Zettlemoyer, L.}
\newblock Bart: Denoising sequence-to-sequence pre-training for natural language generation, translation, and comprehension.
\newblock In {\em ACL\/} (2020).

\bibitem{li2023angle}
{\sc Li, X., and Li, J.}
\newblock Angle-optimized text embeddings.
\newblock {\em arXiv preprint arXiv:2309.12871\/} (2023).

\bibitem{textbooks2}
{\sc Li, Y., Bubeck, S., Eldan, R., Del~Giorno, A., Gunasekar, S., and Lee, Y.~T.}
\newblock Textbooks are all you need ii: \textbf{phi-1.5} technical report.
\newblock {\em arXiv preprint arXiv:2309.05463\/} (2023).

\bibitem{li2021membership}
{\sc Li, Z., and Zhang, Y.}
\newblock Membership leakage in label-only exposures.
\newblock In {\em CCS\/} (2021), pp.~880--895.

\bibitem{lin2004looking}
{\sc Lin, C.-Y., and Och, F.}
\newblock Looking for a few good metrics: Rouge and its evaluation.
\newblock In {\em Ntcir workshop\/} (2004).

\bibitem{liu2022membership}
{\sc Liu, Y., Zhao, Z., Backes, M., and Zhang, Y.}
\newblock Membership inference attacks by exploiting loss trajectory.
\newblock In {\em CCS\/} (2022), pp.~2085--2098.

\bibitem{mattern-etal-2023-membership}
{\sc Mattern, J., Mireshghallah, F., Jin, Z., Schoelkopf, B., Sachan, M., and Berg-Kirkpatrick, T.}
\newblock Membership inference attacks against language models via neighbourhood comparison.
\newblock In {\em Findings of the Association for Computational Linguistics: ACL 2023\/} (Toronto, Canada, July 2023), Association for Computational Linguistics, pp.~11330--11343.

\bibitem{mikolov2010recurrent}
{\sc Mikolov, T., Karafi{\'a}t, M., Burget, L., Cernock{\`y}, J., and Khudanpur, S.}
\newblock Recurrent neural network based language model.
\newblock In {\em Interspeech\/} (2010), vol.~2, Makuhari, pp.~1045--1048.

\bibitem{mireshghallah2022quantifying}
{\sc Mireshghallah, F., Goyal, K., Uniyal, A., Berg-Kirkpatrick, T., and Shokri, R.}
\newblock Quantifying privacy risks of masked language models using membership inference attacks.
\newblock In {\em Proceedings of the 2022 Conference on Empirical Methods in Natural Language Processing\/} (2022), pp.~8332--8347.

\bibitem{mireshghallah2022empirical}
{\sc Mireshghallah, F., Uniyal, A., Wang, T., Evans, D.~K., and Berg-Kirkpatrick, T.}
\newblock An empirical analysis of memorization in fine-tuned autoregressive language models.
\newblock In {\em Proceedings of the 2022 Conference on Empirical Methods in Natural Language Processing\/} (2022), pp.~1816--1826.

\bibitem{nasr2023scalable}
{\sc Nasr, M., Carlini, N., Hayase, J., Jagielski, M., Cooper, A.~F., Ippolito, D., Choquette-Choo, C.~A., Wallace, E., Tram{\`e}r, F., and Lee, K.}
\newblock Scalable extraction of training data from (production) language models.
\newblock {\em arXiv preprint arXiv:2311.17035\/} (2023).

\bibitem{nasr2019comprehensive}
{\sc Nasr, M., Shokri, R., and Houmansadr, A.}
\newblock Comprehensive privacy analysis of deep learning: Passive and active white-box inference attacks against centralized and federated learning.
\newblock In {\em 2019 IEEE symposium on security and privacy (SP)\/} (2019), IEEE, pp.~739--753.

\bibitem{openai2023gpt4}
{\sc OpenAI}.
\newblock Gpt-4 technical report, 2023.

\bibitem{ouyang2022training}
{\sc Ouyang, L., Wu, J., Jiang, X., Almeida, D., Wainwright, C., Mishkin, P., Zhang, C., Agarwal, S., Slama, K., Ray, A., et~al.}
\newblock Training language models to follow instructions with human feedback.
\newblock {\em NeurIPS 35\/} (2022), 27730--27744.

\bibitem{paszke2019pytorch}
{\sc Paszke, A., Gross, S., Massa, F., Lerer, A., Bradbury, J., Chanan, G., Killeen, T., Lin, Z., Gimelshein, N., Antiga, L., et~al.}
\newblock Pytorch: An imperative style, high-performance deep learning library.
\newblock {\em NeurIPS 32\/} (2019).

\bibitem{radford2018improving}
{\sc Radford, A., Narasimhan, K., Salimans, T., Sutskever, I., et~al.}
\newblock Improving language understanding by generative pre-training.

\bibitem{radford2019language}
{\sc Radford, A., Wu, J., Child, R., Luan, D., Amodei, D., Sutskever, I., et~al.}
\newblock Language models are unsupervised multitask learners.
\newblock {\em OpenAI blog 1}, 8 (2019), 9.

\bibitem{reid2024gemini}
{\sc Reid, M., Savinov, N., Teplyashin, D., Lepikhin, D., Lillicrap, T., Alayrac, J.-b., Soricut, R., Lazaridou, A., Firat, O., Schrittwieser, J., et~al.}
\newblock Gemini 1.5: Unlocking multimodal understanding across millions of tokens of context.
\newblock {\em arXiv preprint arXiv:2403.05530\/} (2024).

\bibitem{reimers2019sentence}
{\sc Reimers, N., and Gurevych, I.}
\newblock Sentence-bert: Sentence embeddings using siamese bert-networks.
\newblock {\em arXiv preprint arXiv:1908.10084\/} (2019).

\bibitem{sablayrolles2019white}
{\sc Sablayrolles, A., Douze, M., Schmid, C., Ollivier, Y., and J{\'e}gou, H.}
\newblock White-box vs black-box: Bayes optimal strategies for membership inference.
\newblock In {\em ICML\/} (2019), PMLR, pp.~5558--5567.

\bibitem{salem2018ml}
{\sc Salem, A., Zhang, Y., Humbert, M., Berrang, P., Fritz, M., and Backes, M.}
\newblock Ml-leaks: Model and data independent membership inference attacks and defenses on machine learning models.
\newblock {\em arXiv preprint arXiv:1806.01246\/} (2018).

\bibitem{sankararaman2009genomic}
{\sc Sankararaman, S., Obozinski, G., Jordan, M.~I., and Halperin, E.}
\newblock Genomic privacy and limits of individual detection in a pool.
\newblock {\em Nature genetics 41}, 9 (2009), 965--967.

\bibitem{shao2025explanation}
{\sc Shao, S., Li, Y., Yao, H., He, Y., Qin, Z., and Ren, K.}
\newblock Explanation as a watermark: Towards harmless and multi-bit model ownership verification via watermarking feature attribution.
\newblock In {\em Network and Distributed System Security Symposium\/} (2025).

\bibitem{shi2023detecting}
{\sc Shi, W., Ajith, A., Xia, M., Huang, Y., Liu, D., Blevins, T., Chen, D., and Zettlemoyer, L.}
\newblock Detecting pretraining data from large language models.
\newblock In {\em ICLR\/} (2024).

\bibitem{shokri2017membership}
{\sc Shokri, R., Stronati, M., Song, C., and Shmatikov, V.}
\newblock Membership inference attacks against machine learning models.
\newblock In {\em 2017 IEEE symposium on security and privacy (SP)\/} (2017), IEEE, pp.~3--18.

\bibitem{su2022a}
{\sc Su, Y., Lan, T., Wang, Y., Yogatama, D., Kong, L., and Collier, N.}
\newblock A contrastive framework for neural text generation.
\newblock In {\em NeurIPS\/} (2022), A.~H. Oh, A.~Agarwal, D.~Belgrave, and K.~Cho, Eds.

\bibitem{tanay2016boundary}
{\sc Tanay, T., and Griffin, L.}
\newblock A boundary tilting persepective on the phenomenon of adversarial examples.
\newblock {\em arXiv preprint arXiv:1608.07690\/} (2016).

\bibitem{tang2023assessing}
{\sc Tang, R., Lueck, G., Quispe, R., Inan, H., Kulkarni, J., and Hu, X.}
\newblock Assessing privacy risks in language models: A case study on summarization tasks.
\newblock In {\em Findings of the Association for Computational Linguistics: EMNLP 2023\/} (2023), pp.~15406--15418.

\bibitem{team2023gemini}
{\sc Team, G., Anil, R., Borgeaud, S., Alayrac, J.-B., Yu, J., Soricut, R., Schalkwyk, J., Dai, A.~M., Hauth, A., Millican, K., et~al.}
\newblock Gemini: a family of highly capable multimodal models.
\newblock {\em arXiv preprint arXiv:2312.11805\/} (2023).

\bibitem{team2024gemma}
{\sc Team, G., Riviere, M., Pathak, S., Sessa, P.~G., Hardin, C., Bhupatiraju, S., Hussenot, L., Mesnard, T., Shahriari, B., Ram{\'e}, A., et~al.}
\newblock Gemma 2: Improving open language models at a practical size.
\newblock {\em arXiv preprint arXiv:2408.00118\/} (2024).

\bibitem{tirumala2022memorization}
{\sc Tirumala, K., Markosyan, A., Zettlemoyer, L., and Aghajanyan, A.}
\newblock Memorization without overfitting: Analyzing the training dynamics of large language models.
\newblock {\em NeurIPS 35\/} (2022), 38274--38290.

\bibitem{touvron2023llamaa}
{\sc Touvron, H., Lavril, T., Izacard, G., Martinet, X., Lachaux, M.-A., Lacroix, T., Rozi{\`e}re, B., Goyal, N., Hambro, E., Azhar, F., et~al.}
\newblock Llama: Open and efficient foundation language models.
\newblock {\em arXiv preprint arXiv:2302.13971\/} (2023).

\bibitem{touvron2023llamab}
{\sc Touvron, H., Martin, L., Stone, K., Albert, P., Almahairi, A., Babaei, Y., Bashlykov, N., Batra, S., Bhargava, P., Bhosale, S., et~al.}
\newblock Llama 2: Open foundation and fine-tuned chat models.
\newblock {\em arXiv preprint arXiv:2307.09288\/} (2023).

\bibitem{vaithilingam2022expectation}
{\sc Vaithilingam, P., Zhang, T., and Glassman, E.~L.}
\newblock Expectation vs. experience: Evaluating the usability of code generation tools powered by large language models.
\newblock In {\em Chi conference on human factors in computing systems extended abstracts\/} (2022), pp.~1--7.

\bibitem{wang2021gpt}
{\sc Wang, B., and Komatsuzaki, A.}
\newblock Gpt-j-6b: A 6 billion parameter autoregressive language model, 2021.

\bibitem{watson2021importance}
{\sc Watson, L., Guo, C., Cormode, G., and Sablayrolles, A.}
\newblock On the importance of difficulty calibration in membership inference attacks.
\newblock In {\em ICLR\/} (2021).

\bibitem{wolf2020transformers}
{\sc Wolf, T., Debut, L., Sanh, V., Chaumond, J., Delangue, C., Moi, A., Cistac, P., Rault, T., Louf, R., Funtowicz, M., et~al.}
\newblock Transformers: State-of-the-art natural language processing.
\newblock In {\em Proceedings of the 2020 conference on empirical methods in natural language processing: system demonstrations\/} (2020), pp.~38--45.

\bibitem{wu2024you}
{\sc WU, Y., Qiu, H., Guo, S., Li, J., and Zhang, T.}
\newblock You only query once: An efficient label-only membership inference attack.
\newblock In {\em The Twelfth International Conference on Learning Representations\/} (2024).

\bibitem{bge_embedding}
{\sc Xiao, S., Liu, Z., Zhang, P., and Muennighoff, N.}
\newblock C-pack: Packaged resources to advance general chinese embedding, 2023.

\bibitem{xie2024recall}
{\sc Xie, R., Wang, J., Huang, R., Zhang, M., Ge, R., Pei, J., Gong, N.~Z., and Dhingra, B.}
\newblock Recall: Membership inference via relative conditional log-likelihoods.
\newblock {\em arXiv preprint arXiv:2406.15968\/} (2024).

\bibitem{yao2024machine}
{\sc Yao, J., Chien, E., Du, M., Niu, X., Wang, T., Cheng, Z., and Yue, X.}
\newblock Machine unlearning of pre-trained large language models.
\newblock {\em arXiv preprint arXiv:2402.15159\/} (2024).

\bibitem{ye2022enhanced}
{\sc Ye, J., Maddi, A., Murakonda, S.~K., Bindschaedler, V., and Shokri, R.}
\newblock Enhanced membership inference attacks against machine learning models.
\newblock In {\em CCS\/} (2022), pp.~3093--3106.

\bibitem{yeom2018privacy}
{\sc Yeom, S., Giacomelli, I., Fredrikson, M., and Jha, S.}
\newblock Privacy risk in machine learning: Analyzing the connection to overfitting.
\newblock In {\em 2018 IEEE 31st computer security foundations symposium (CSF)\/} (2018), IEEE, pp.~268--282.

\bibitem{zhang2024min}
{\sc Zhang, J., Sun, J., Yeats, E., Ouyang, Y., Kuo, M., Zhang, J., Yang, H., and Li, H.}
\newblock Min-k\%++: Improved baseline for detecting pre-training data from large language models.
\newblock {\em arXiv preprint arXiv:2404.02936\/} (2024).

\bibitem{zhang2022opt}
{\sc Zhang, S., Roller, S., Goyal, N., Artetxe, M., Chen, M., Chen, S., Dewan, C., Diab, M., Li, X., Lin, X.~V., Mihaylov, T., Ott, M., Shleifer, S., Shuster, K., Simig, D., Koura, P.~S., Sridhar, A., Wang, T., and Zettlemoyer, L.}
\newblock Opt: Open pre-trained transformer language models, 2022.

\end{thebibliography}

\clearpage
\onecolumn
\appendix


\section{Rationale behind Our Case Studies Setup}
\label{sec: Case_study_setup}
The rationale behind our experimental setup is threefold: 1) As a standard source of high-quality text, Wikipedia is a commonly used pre-training data source \cite{shi2023detecting}; 2) The predecessors of Gemini-1.5 and GPT-3.5 (i.e., the GPT-3 series \cite{brown2020language} and PaLM \cite{chowdhery2023palm}) have both been explicitly reported to be pre-trained on Wikipedia; 3) The knowledge cutoff dates for both Gemini-1.5 and GPT-3.5 precede March 2024. In summary, despite the lack of detailed knowledge about the training data for these two commercial LLMs, our case study is highly likely to align with real-world scenarios. Therefore, despite the lack of detailed knowledge about the training data for these two commercial LLMs, our case studies are highly likely to align with real-world scenarios.

\section{Balanced Accuracy Results on MIMIR}
\label{sec: main Acc results}

Table \ref{table:main results acc on MIMIR} demonstrates that PETAL not only substantially outperforms prior label-only attacks but also achieves comparable performance to existing logits-based attacks in terms of balanced accuracy. This is expected as both balanced accuracy and AUC metrics reflect the average level of privacy leakage, and thus typically exhibit similar trends.

\begin{table*}[!t]
\centering
\setlength{\tabcolsep}{4.0pt}
\caption{Balanced accuracy results of various attacks on different subsets of the Pile. We bold our results when they surpass all label-only attacks and underline them when they are comparable to logits-based attacks (i.e., surpassing at least two logits-based attacks).}
\scalebox{0.91}
{
\begin{tabular}{l|cccccccccccccccc}
\toprule
Subsets & \multicolumn{4}{c}{arXiv}&  \multicolumn{4}{c}{DM Mathematics}& \multicolumn{4}{c}{GitHub}& \multicolumn{4}{c}{HackerNews}\\
\cmidrule(l{5pt}r{5pt}){2-5}\cmidrule(l{5pt}r{5pt}){6-9}\cmidrule(l{5pt}r{5pt}){10-13}\cmidrule(l{5pt}r{5pt}){14-17}
Models & 160M& 1.4B& 2.8B& 6.9B& 160M& 1.4B& 2.8B& 6.9B& 160M& 1.4B& 2.8B& 6.9B& 160M& 1.4B& 2.8B& 6.9B\\
\midrule
\textbf{\textit{Logits-based Attacks}}&&&&&&&&&&&&\\
PPL attack&0.59&0.63&0.62&0.61&0.83&0.83&0.81&0.79&0.81&0.82&0.82&0.83&0.59&0.58&0.59&0.58\\
reference attack&0.61&0.63&0.63&0.62&0.80&0.78&0.76&0.75&0.61&0.63&0.63&0.64&0.53&0.54&0.54&0.54\\
zlib attack&0.61&0.64&0.63&0.64&0.77&0.75&0.76&0.72&0.78&0.80&0.81&0.82&0.57&0.58&0.59&0.58\\
neighborhood attack&0.64&0.66&0.65&0.66&0.67&0.70&0.69&0.71&0.75&0.77&0.77&0.78&0.55&0.54&0.55&0.54\\
MIN-K\% PROB&0.59&0.61&0.61&0.60&0.66&0.66&0.64&0.64&0.77&0.80&0.81&0.81&0.57&0.56&0.58&0.58\\
\textbf{\textit{label-only attacks}}&&&&&&&&&&&&\\
WS attack&0.52&0.52&0.54&0.52&0.53&0.51&0.54&0.51&0.68&0.72&0.73&0.72&0.52&0.52&0.53&0.52\\
RS attack&0.52&0.53&0.52&0.54&0.50&0.50&0.51&0.51&0.71&0.73&0.73&0.73&0.52&0.51&0.53&0.52\\
BT attack&0.54&0.52&0.54&0.55&0.52&0.52&0.52&0.52&0.68&0.70&0.70&0.71&0.52&0.52&0.53&0.54\\
PETAL (Ours)&\textbf{0.54}&\textbf{0.57}&\textbf{0.56}&\textbf{0.56}&\ul{\textbf{0.82}}&\ul{\textbf{0.83}}&\ul{\textbf{0.82}}&\ul{\textbf{0.80}}&\ul{\textbf{0.77}}&\ul{\textbf{0.81}}&\ul{\textbf{0.81}}&\ul{\textbf{0.82}}&\ul{\textbf{0.59}}&\ul{\textbf{0.58}}&\ul{\textbf{0.59}}&\ul{\textbf{0.59}}\\
\vspace{-.9em}\\
\toprule
Subsets & \multicolumn{4}{c}{Pile CC}&  \multicolumn{4}{c}{PubMed Central}& \multicolumn{4}{c}{Wikipedia}& \multicolumn{4}{c}{Average}\\
\cmidrule(l{5pt}r{5pt}){2-5}\cmidrule(l{5pt}r{5pt}){6-9}\cmidrule(l{5pt}r{5pt}){10-13}\cmidrule(l{5pt}r{5pt}){14-17}
Models & 160M& 1.4B& 2.8B& 6.9B& 160M& 1.4B& 2.8B& 6.9B& 160M& 1.4B& 2.8B& 6.9B& 160M& 1.4B& 2.8B& 6.9B\\
\midrule
\textbf{\textit{Logits-based Attacks}}&&&&&&&&&&&&\\
PPL attack&0.53&0.54&0.56&0.55&0.67&0.64&0.63&0.65&0.60&0.60&0.60&0.60&0.66&0.66&0.66&0.66\\
reference attack&0.52&0.53&0.52&0.53&0.64&0.64&0.63&0.64&0.55&0.58&0.59&0.60&0.61&0.62&0.61&0.62\\
zlib attack&0.51&0.52&0.53&0.54&0.66&0.65&0.65&0.67&0.57&0.58&0.59&0.59&0.64&0.65&0.65&0.65\\
neighborhood attack&0.54&0.54&0.53&0.54&0.65&0.65&0.64&0.66&0.54&0.55&0.55&0.56&0.62&0.63&0.63&0.63\\
MIN-K\% PROB&0.54&0.53&0.54&0.53&0.59&0.56&0.56&0.57&0.58&0.58&0.59&0.60&0.61&0.61&0.62&0.62\\
\textbf{\textit{label-only attacks}}&&&&&&&&&&&&\\
WS attack&0.52&0.53&0.52&0.53&0.52&0.53&0.55&0.55&0.52&0.54&0.54&0.54&0.55&0.55&0.56&0.55\\
RS attack&0.54&0.53&0.52&0.53&0.54&0.55&0.55&0.55&0.54&0.54&0.55&0.55&0.55&0.55&0.56&0.56\\
BT attack&0.52&0.53&0.52&0.52&0.54&0.53&0.53&0.56&0.53&0.54&0.55&0.53&0.55&0.55&0.55&0.56\\
PETAL (Ours)&\ul{\textbf{0.54}}&\ul{\textbf{0.55}}&\ul{\textbf{0.55}}&\ul{\textbf{0.55}}&\textbf{0.63}&\textbf{0.63}&\ul{\textbf{0.63}}&\textbf{0.62}&\ul{\textbf{0.59}}&\ul{\textbf{0.61}}&\ul{\textbf{0.59}}&\ul{\textbf{0.60}}&\ul{\textbf{0.64}}&\ul{\textbf{0.65}}&\ul{\textbf{0.65}}&\ul{\textbf{0.65}}\\
\bottomrule
\end{tabular}
}
\label{table:main results acc on MIMIR}
\end{table*}

\clearpage

\section{ROC Curves of Main Experiments}
\label{sec: roc curves}

\begin{figure*}[!t]
    \centering
    \subfloat[WikiMIA]{\includegraphics[width=1.75in]{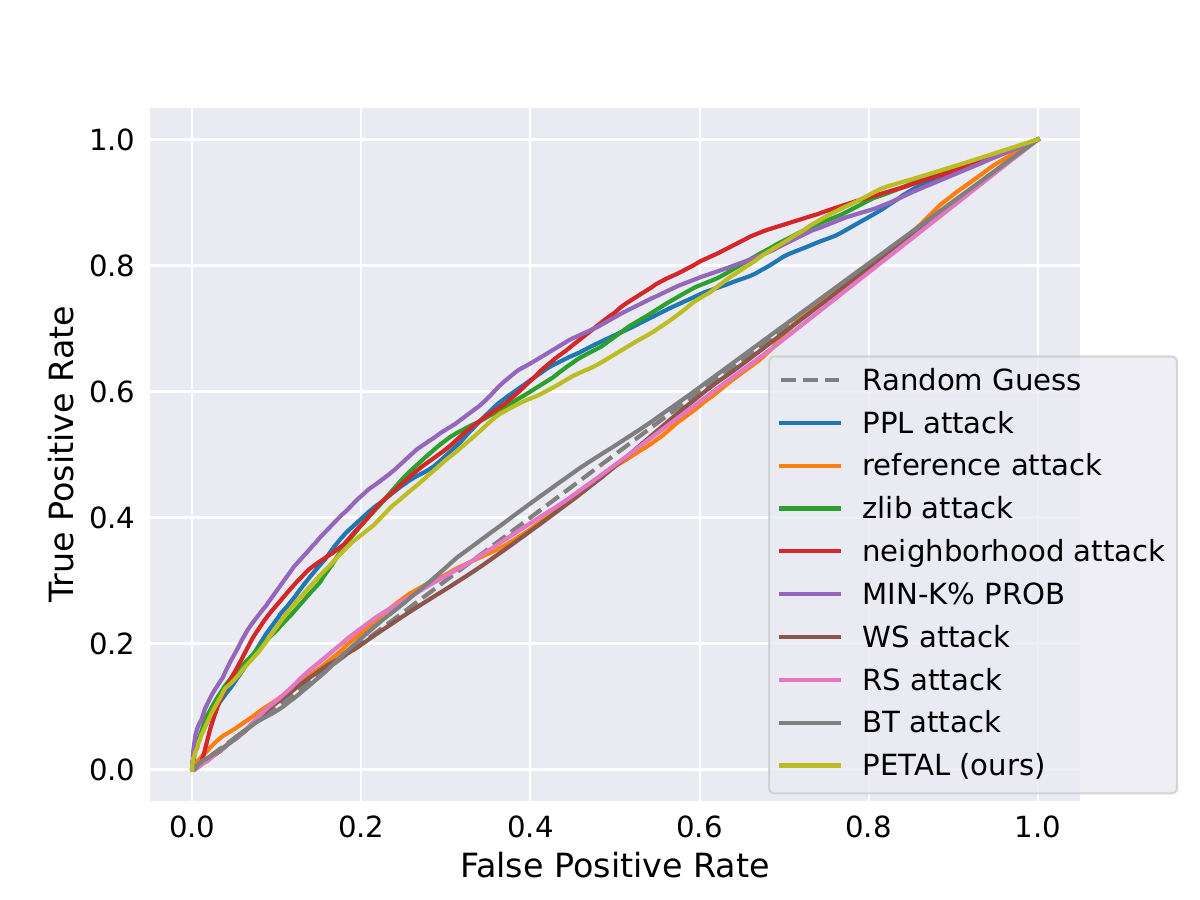} }
    \subfloat[DM Mathematics]{\includegraphics[width=1.75in]{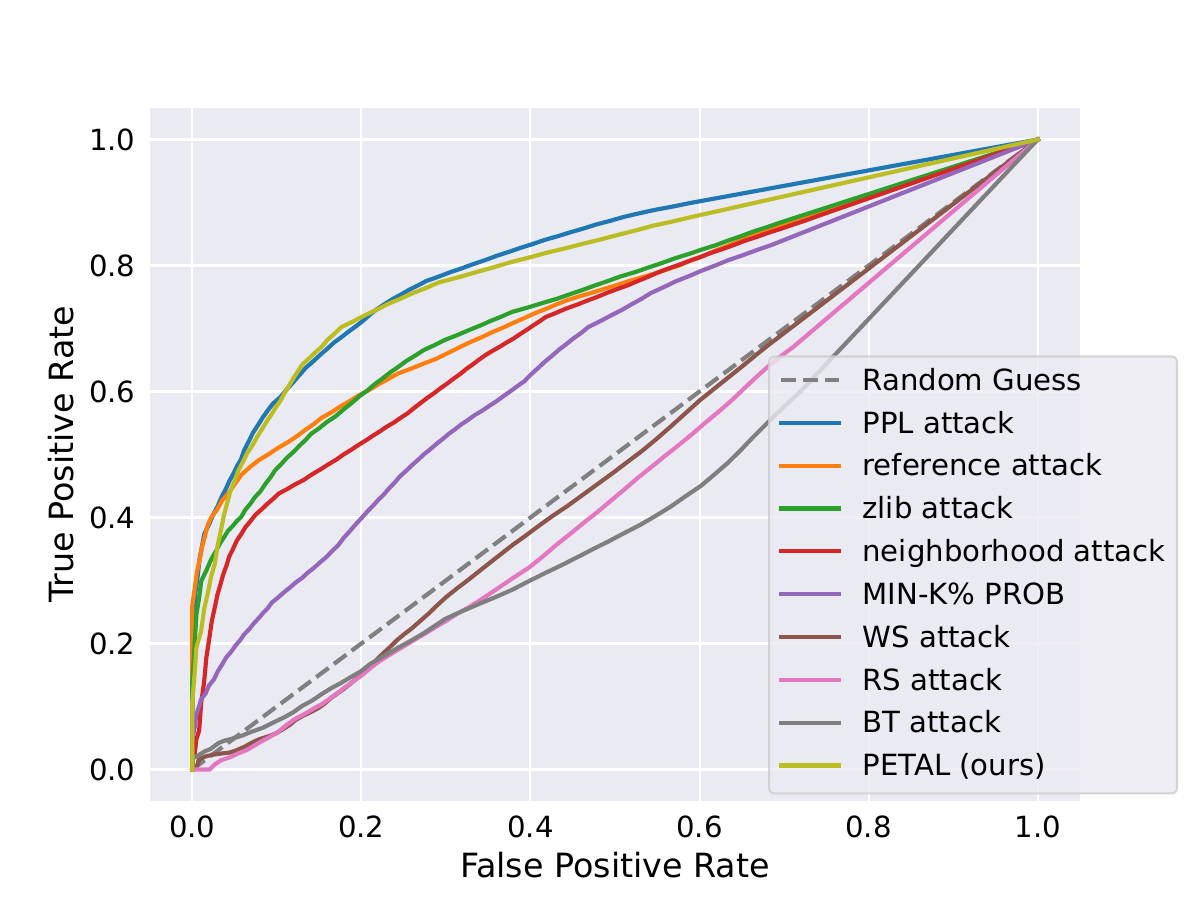} }
    \subfloat[arXiv]{\includegraphics[width=1.75in]{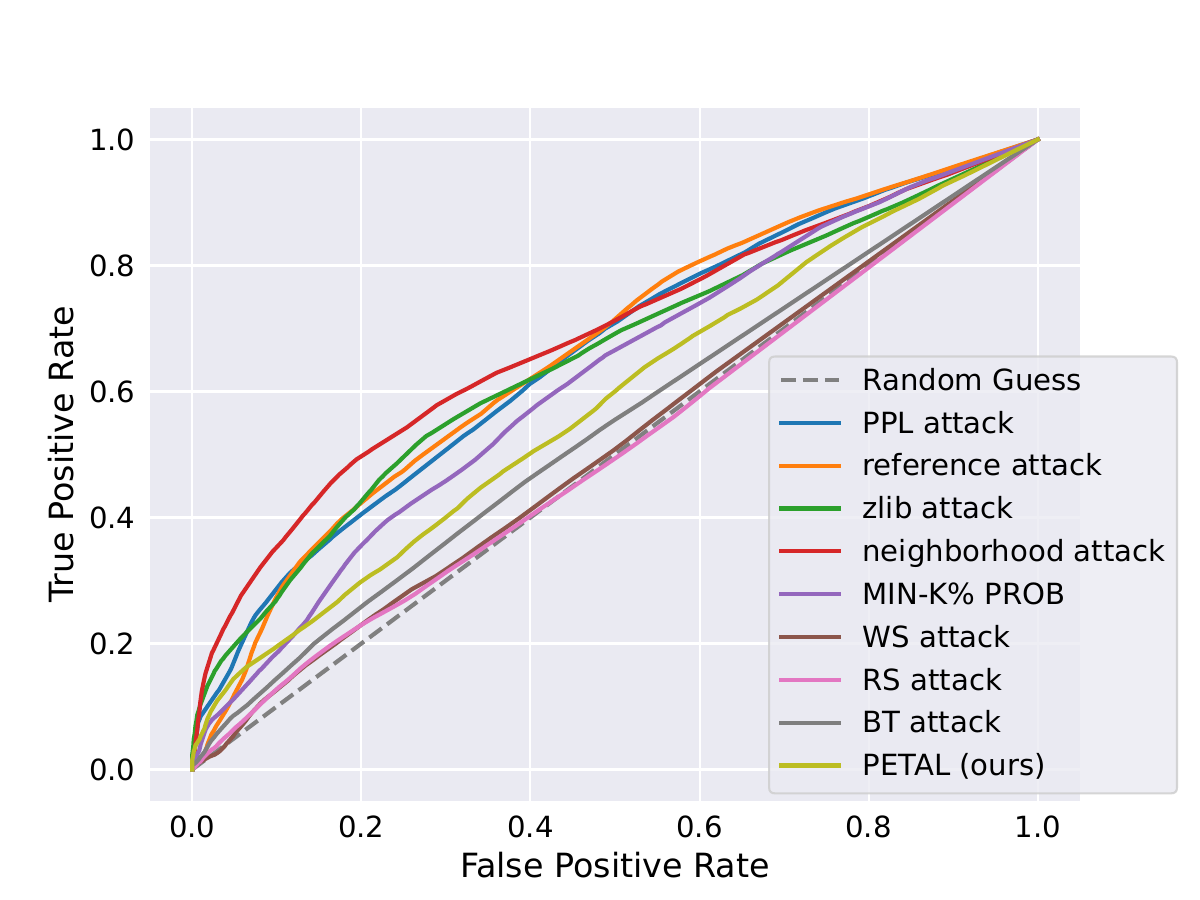} }
    \subfloat[GitHub]{\includegraphics[width=1.75in]{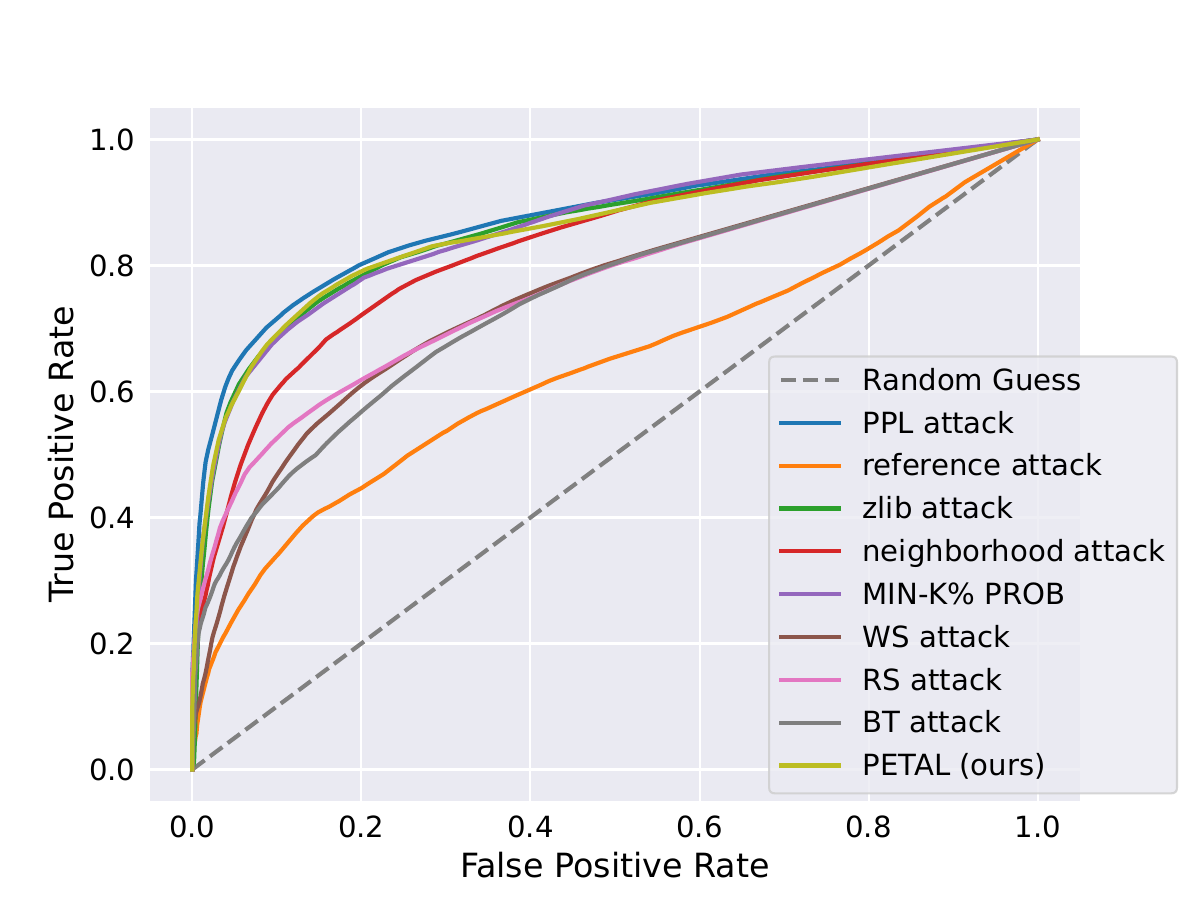} }
    \\
    \subfloat[Hackerbews]{\includegraphics[width=1.75in]{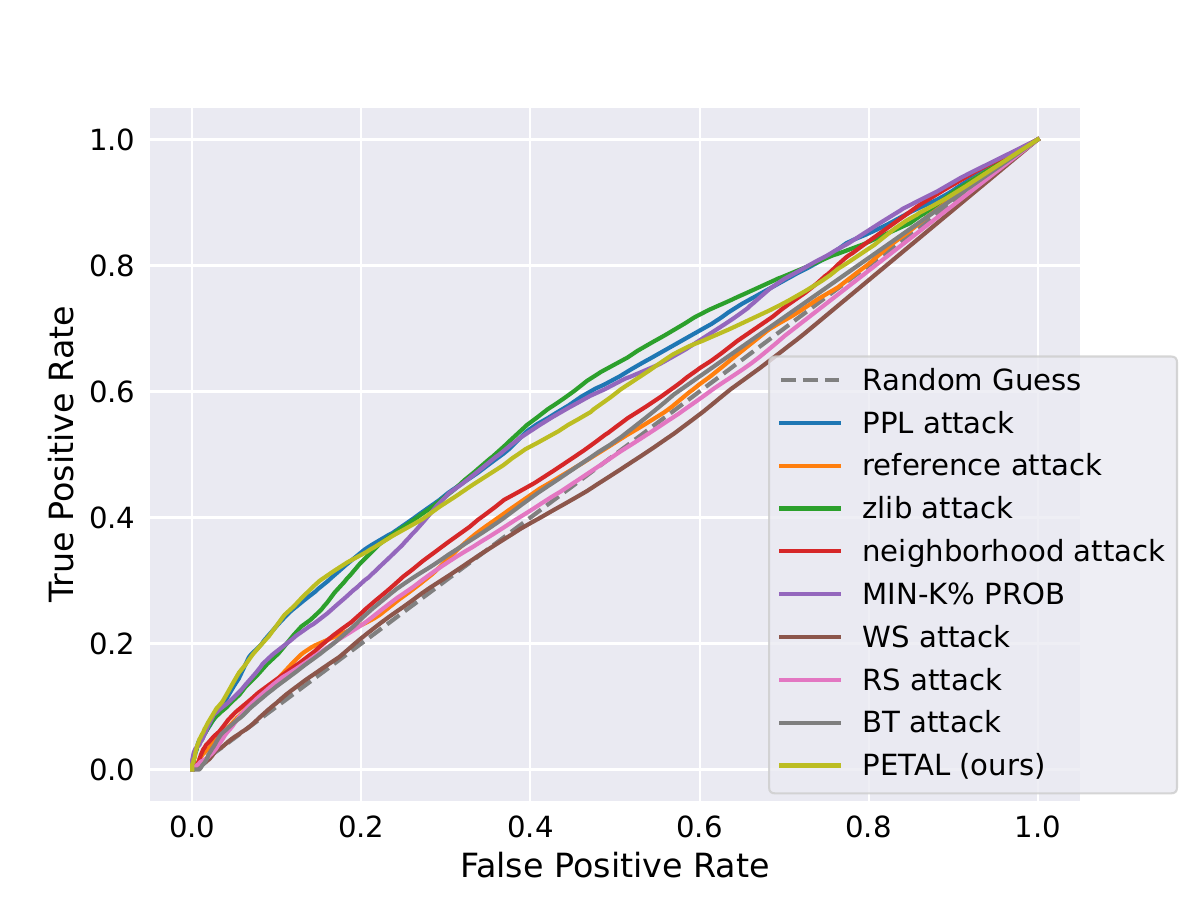} }
    \subfloat[Pile CC]{\includegraphics[width=1.75in]{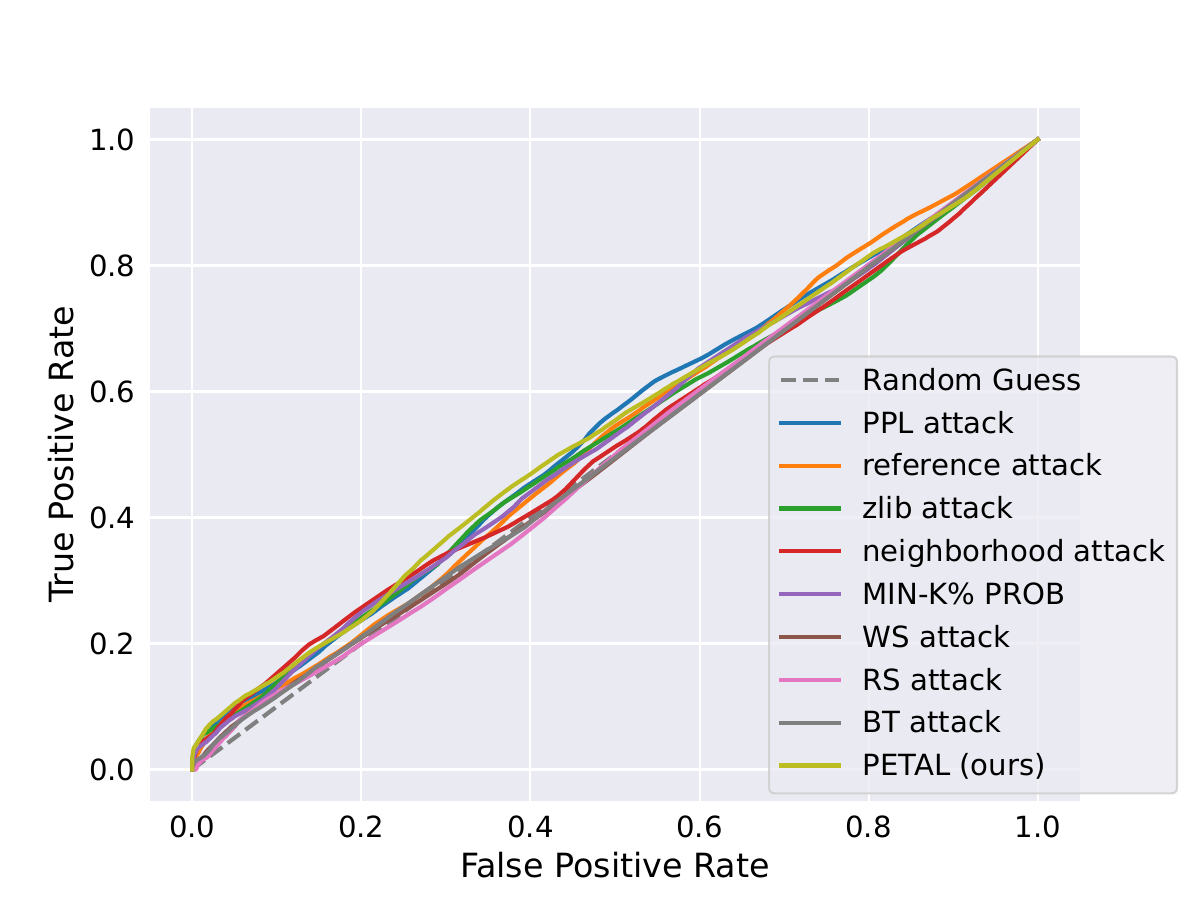} }
    \subfloat[PubMed Central]{\includegraphics[width=1.75in]{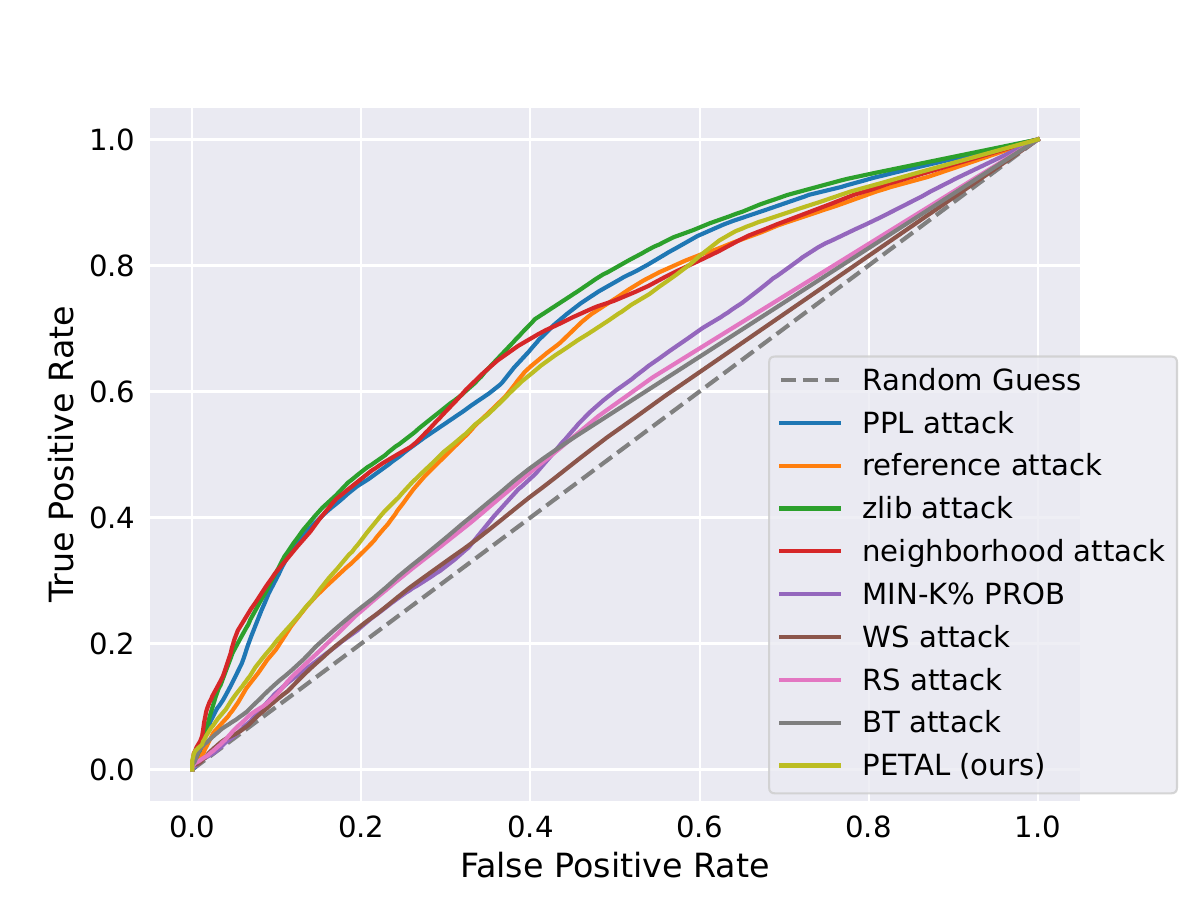} }
    \subfloat[Wikipedia]{\includegraphics[width=1.75in]{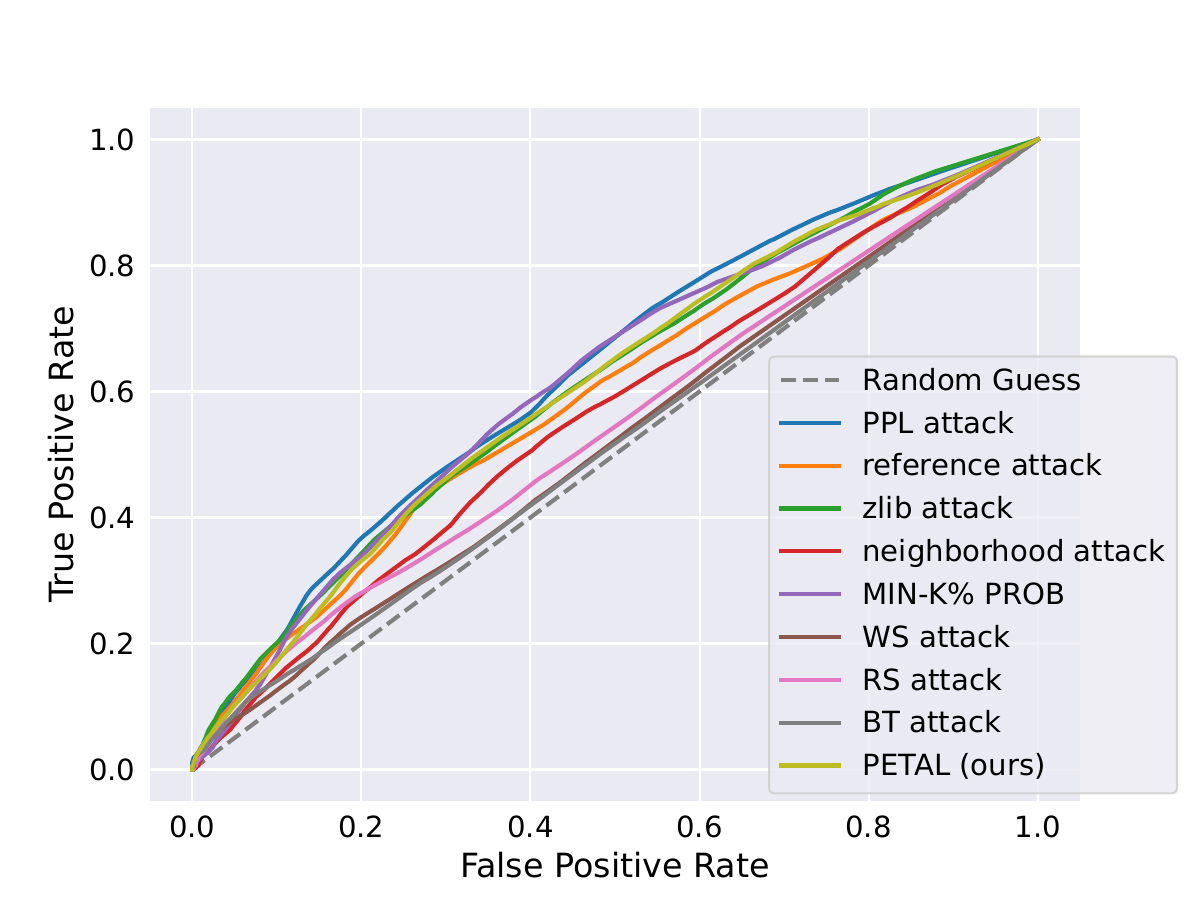} }
    \caption{The ROC curves show attack results on the WikiMIA benchmark and seven subsets of the MIMIR benchmark, with the target model set as Pythia-6.9B.}
\label{fig: roc curves}
\end{figure*}

Figure \ref{fig: roc curves} displays the complete ROC curves for attack results on the WikiMIA benchmark and seven subsets of the MIMIR benchmark. As shown, PETAL generally achieves TPRs comparable to those of logits-based attacks across varying FPR settings.

\section{Additional Experimental Results on Different Estimation Granularity}
\label{sec: estimation granularity}
\begin{figure}[h]
    \centering
    \includegraphics[width=0.47\textwidth]{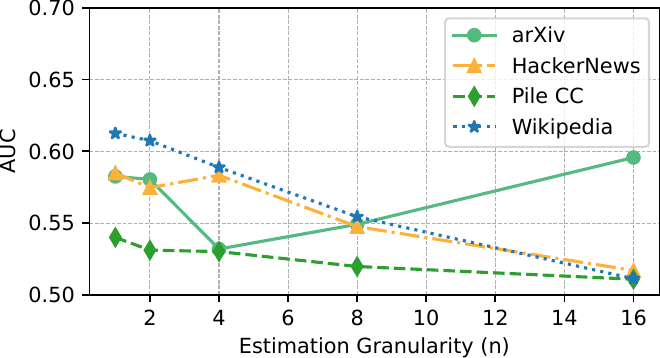}
    \caption{Additional AUC values of our PETAL against Pythia-6.9B. The granularity of probability estimation is set as 1, 2, 4, 8, and 16.}
    \label{Figure: additional ablation on estimation granularity}
\end{figure}

Figure \ref{Figure: additional ablation on estimation granularity} shows the attack results on the other four subsets of MIMIR under different estimation granularity settings. It is worth noting that in contrast to WikiMIA and the other six subsets of MIMIR, the attack results on the arXiv subset exhibit an anomalous increase as the estimation granularity rises from 4 to 16. This unexpected behavior is potentially related to the special distribution characteristics of the arXiv dataset. We will leave this as an open problem and address it in our future work.

\clearpage

\section{Balanced Accuracy Results on Paraphrased WikiMIA}
\label{sec: balanced acc on wiki-para}
The results in Table \ref{table:as_paraphrasing_additional} demonstrate that PETAL's balanced accuracy, similar to its AUC performance, shows little to no decline when input text is paraphrased. This is expected as semantic information contained in tokens is hardly altered by text paraphrasing.

\begin{table*}[!t]
\centering
\setlength{\tabcolsep}{4.0pt}
\caption{Balanced accuracy results of various attacks on WikiMIA when the input text is paraphrased. Values in parentheses indicate the difference in attack performance between paraphrased and non-paraphrased text.}
\scalebox{1.0}
{
\begin{tabular}{l|cccc}
\toprule
Metrics & \multicolumn{4}{c}{Balanced Accuracy $\uparrow$} \\
\cmidrule(l{5pt}r{5pt}){2-5}
Models & Pythia-6.9B& OPT-6.7B& Falcon-7B& LLaMA2-13B\\
\midrule
\textbf{\textit{Logits-based Attacks}}&&&&\\
PPL attack&0.61 (-0.00)&0.58 (+0.00)&0.55 (-0.01)&0.58 (-0.00)\\
reference attack&0.52 (+0.01)&0.50 (+0.00)&0.50 (+0.00)&0.52 (+0.00)\\
zlib attack&0.61 (-0.01)&0.59 (+0.00)&0.56 (+0.00)&0.58 (-0.00)\\
neighborhood attack&0.62 (+0.00)&0.61 (-0.01)&0.56 (-0.01)&0.60 (+0.00)\\
MIN-K\% PROB&0.62 (-0.01)&0.60 (-0.01)&0.53 (+0.00)&0.55 (-0.01)\\
\textbf{\textit{Label-only Attacks}}&&&&\\
PETAL (Ours)&0.61 (+0.00)&0.59 (-0.00)&0.57 (+0.00)&0.58 (+0.00)\\
\bottomrule
\end{tabular}
}
\vspace{-0.5em}
\label{table:as_paraphrasing_additional}
\end{table*}

\section{Additional Experimental Results on Training Data Deduplication}
\label{sec: additional results on granularity}
As shown in Table \ref{table:additional_as_deduplicaion_acc} \& \ref{table:additional_as_deduplicaion_tpr}, data deduplication only slightly reduces the balanced accuracy of MIAs (to an almost negligible extent), and does not affect the TPR at 1\% FPR at all. These results further indicate that coarse-grained deduplication is ineffective in mitigating the privacy threats posed by MIAs.

\begin{table*}[h]
\centering
\setlength{\tabcolsep}{4.0pt}
\caption{Balanced accuracy of attacks on two subsets of MIMIR when the target model is pre-trained on deduplicated training data. Values in parentheses denote the difference in attack performance between deduped and non-deduped target models.}
\scalebox{0.88}
{
\begin{tabular}{l|cccccccc}
\toprule
Subsets & \multicolumn{4}{c}{GitHub}&  \multicolumn{4}{c}{HackerNews}\\
\cmidrule(l{5pt}r{5pt}){2-5}\cmidrule(l{5pt}r{5pt}){6-9}
Models & 160M& 1.4B& 2.8B& 6.9B& 160M& 1.4B& 2.8B& 6.9B\\
\midrule
\textbf{\textit{Logits-based Attacks}}&&&&&&&&\\
PPL attack&0.79 (-0.01)&0.81 (-0.00)&0.82 (+0.00)&0.82 (-0.01)&0.58 (-0.01)&0.58 (-0.00)&0.58 (-0.01)&0.58 (-0.00)\\
reference attack&0.63 (+0.02)&0.62 (-0.01)&0.62 (-0.01)&0.62 (-0.02)&0.55 (+0.02)&0.53 (-0.01)&0.54 (+0.01)&0.53 (-0.00)\\
zlib attack&0.78 (-0.00)&0.80 (+0.00)&0.81 (+0.00)&0.81 (-0.01)&0.58 (+0.00)&0.58 (-0.00)&0.59 (-0.00)&0.58 (-0.01)\\
neighborhood attack&0.77 (+0.02)&0.77 (-0.00)&0.77 (+0.00)&0.76 (-0.02)&0.55 (-0.00)&0.54 (-0.00)&0.54 (-0.00)&0.54 (+0.01)\\
MIN-K\% PROB&0.75 (-0.02)&0.80 (+0.01)&0.80 (-0.00)&0.81 (-0.01)&0.56 (-0.01)&0.56 (+0.00)&0.57 (-0.01)&0.56 (-0.02)\\
\textbf{\textit{Label-only Attacks}}&&&&&&&&\\
PETAL (Ours)&0.78 (+0.01)&0.81 (-0.00)&0.81 (-0.00)&0.81 (-0.01)&0.58 (-0.00)&0.58 (-0.00)&0.59 (-0.00)&0.59 (-0.00)\\
\bottomrule
\end{tabular}
}
\vspace{-0.5em}
\label{table:additional_as_deduplicaion_acc}
\end{table*}

\begin{table*}[!h]
\centering
\setlength{\tabcolsep}{4.0pt}
\caption{The TPR results of attacks at 1\% FPR on two subsets of MIMIR when the target model is pre-trained on deduplicated training data. Values in parentheses denote the difference in performance between deduped and non-deduped target models.}
\scalebox{0.78}
{
\begin{tabular}{l|cccccccc}
\toprule
Subsets & \multicolumn{4}{c}{GitHub}&  \multicolumn{4}{c}{HackerNews}\\
\cmidrule(l{5pt}r{5pt}){2-5}\cmidrule(l{5pt}r{5pt}){6-9}
Models & 160M& 1.4B& 2.8B& 6.9B& 160M& 1.4B& 2.8B& 6.9B\\
\midrule
\textbf{\textit{Logits-based Attacks}}&&&&&&&&\\
PPL attack&30.0\% (-0.4\%)&36.8\% (+1.2\%)&44.4\% (+1.2\%)&52.0\% (+3.2\%)&4.0\% (-1.2\%)&4.8\% (+0.0\%)&6.0\% (+0.0\%)&5.2\% (-0.4\%)\\
reference attack&9.6\% (+0.8\%)&12.4\% (-2.8\%)&11.2\% (-1.6\%)&10.0\% (+0.4\%)&3.2\% (+0.8\%)&0.4\% (-0.8\%)&1.2\% (+0.0\%)&1.6\% (+0.4\%)\\
zlib attack&17.6\% (+0.4\%)&29.6\% (+2.0\%)&30.0\% (-1.2\%)&40.0\% (+3.2\%)&4.8\% (+0.0\%)&4.4\% (-0.4\%)&4.4\% (+0.0\%)&5.2\% (+0.0\%)\\
neighborhood attack&25.2\% (+2.8\%)&24.0\% (-8.0\%)&28.4\% (+2.4\%)&28.4\% (+1.2\%)&2.8\% (+0.4\%)&6.0\% (+5.2\%)&3.6\% (+1.2\%)&3.2\% (+2.0\%)\\
MIN-K\% PROB&19.2\% (-7.2\%)&25.2\% (+1.6\%)&30.8\% (-0.8\%)&29.2\% (+0.8\%)&1.6\% (-0.4\%)&2.4\% (-0.8\%)&3.6\% (-1.2\%)&4.4\% (-0.4\%)\\
\textbf{\textit{Label-only Attacks}}&&&&&&&&\\
PETAL (Ours)&30.8\% (+4.4\%)&40.4\% (+9.2\%)&38.0\% (+4.4\%)&40.8\% (-3.6\%)&7.6\% (-0.8\%)&4.4\% (+0.8\%)&5.2\% (+0.8\%)&6.0\% (-0.8\%)\\
\bottomrule
\end{tabular}
}
\label{table:additional_as_deduplicaion_tpr}
\end{table*}
\end{document}